\documentclass[sigconf,9pt]{acmart}

\usepackage{amsmath,amssymb,amsfonts}
\usepackage{textgreek}
\usepackage{gensymb}
\usepackage{array}
\usepackage{algorithm}
\usepackage{algpseudocode}
\usepackage{verbatim}
\usepackage{enumitem}
\usepackage{times}
\usepackage{hyperref}
\usepackage{graphicx}
\usepackage{textgreek}
\usepackage{gensymb}
\usepackage{array}
\usepackage{algorithm}
\usepackage{algpseudocode}
\usepackage{enumitem}
\usepackage{nccmath}
\usepackage{comment}
\usepackage{subcaption}
\usepackage{booktabs}
\usepackage{diagbox}
\usepackage{amsmath}
\usepackage{cleveref}
\usepackage{lipsum}
\usepackage{makecell}
\usepackage{filecontents}

\def\BibTeX{{\rm B\kern-.05em{\sc i\kern-.025em b}\kern-.08emT\kern-.1667em\lower.7ex\hbox{E}\kern-.125emX}}
    
\renewcommand\footnotetextcopyrightpermission[1]{}
\copyrightyear{2019}
\acmYear{2019}
\setcopyright{none}
\acmConference[ICS '19]{ICS '19: International Conference on Supercomputing}{June 26--28, 2019}{Phoenix, AZ}
\acmBooktitle{ICS '19: International Conference on Supercomputing, June 26--28, 2019, Phoenix, AZ}

\begin{document}

%
\title[Fully Integrated On-FPGA Molecular Dynamics Simulations]{Fully Integrated On-FPGA Molecular Dynamics Simulations}

%

\author{Chen Yang}
\affiliation{\institution{Boston University}}
\email{cyang90@bu.edu}

\author{Tong Geng}
\affiliation{\institution{Boston University}}
\email{tgeng@bu.edu}

\author{Tianqi Wang}
\affiliation{\institution{University of Science and Technology of China}}
\email{tqwang@mail.ustc.edu.cn}

\author{Rushi Patel}
\affiliation{\institution{Boston University}}
\email{ruship@bu.edu}

\author{Qingqing Xiong}
\affiliation{\institution{Boston University}}
\email{qx@bu.edu}

\author{Ahmed Sanaullah}
\affiliation{\institution{Boston University}}
\email{sanaulah@bu.edu}

\author{Jiayi Sheng}
\affiliation{\institution{Falcon Computing Solutions, Inc.}}
\email{jysheng@falcon-computing.com}

\author{Charles Lin}
\affiliation{\institution{Silicon Therapeutics}}
\email{charles.lin@silicontx.com}

\author{Vipin Sachdeva}
\affiliation{\institution{Silicon Therapeutics}}
\email{vipin@silicontx.com}

\author{Woody Sherman}
\affiliation{\institution{Silicon Therapeutics}}
\email{woody@silicontx.com}

\author{Martin C. Herbordt}
\affiliation{\institution{Boston University}}
\email{herbordt@bu.edu}


\renewcommand{\shortauthors}{Author 1 and Author 2, et al.}

%
\begin{abstract}


The implementation of Molecular Dynamics (MD) on FPGAs has received substantial attention. Previous work, however, has consisted of either proof-of-concept implementations of components, usually the range-limited force; full systems, but with much of the work shared by the host CPU; or prototype demonstrations, e.g., using OpenCL, that neither implement a whole system nor have competitive performance. In this paper, we present what we believe to be the first full-scale FPGA-based simulation engine, and show that its performance is competitive with a GPU (running Amber in an industrial production environment). The system features on-chip particle data storage and management, short- and long-range force evaluation, as well as bonded forces, motion update, and particle migration. Other contributions of this work include exploring numerous architectural trade-offs and analysis on various mappings schemes among particles/cells and the various on-chip compute units. The potential impact is that this system promises to be the basis for long timescale Molecular Dynamics with a commodity cluster.

\end{abstract}


%
%
%

%
\maketitle

\section{Introduction}
\label{sec:intro}


There are dozens of MD packages in production use (e.g.,  \cite{Case18,Phillips05,vanderSpoel05,Eastman10,Plimpton95}), many of which have been successfully accelerated with GPUs. Scaling, however, remains problematic for the small simulations (20K-50K particles) commonly used in critical applications, e.g., drug design~\cite{Cournia17,NVIDIA17}, where long timescales are also extremely beneficial. 
Simulation of long timescales of small molecules is, of course, a motivation for the Anton family of ASIC-based MD engines \cite{Shaw07,Grossman15}. Anton addresses scalability by having direct communication links--application layer to application layer--among the integrated circuits (ICs) in the cluster. But while ASIC-based solutions can have orders-of-magnitude better performance than commodity clusters, they may also have issues with general availability, plus all the problems inherent with small-run ASIC-based systems.

FPGAs have been explored as possible MD accelerators for many years 
\cite{Azizi04,Gu05b,Hamada05,Scrofano06b,Kindratenko06,Alam07,Cong16}. The first generation of complete FPGA/MD systems accelerated only the range limited (RL) force and used CPUs for the rest of the computation. While performance was sometimes competitive, high cost and lack of availability of FPGA systems meant that they were never in production use. In the last few years, however, it has been shown that FPGA clusters can have performance approaching that of ASIC clusters for the Long Range force computation (LR) \cite{Sheng14,Sheng15b,Lawande16,Sheng17}, the part of MD that is most difficult to scale. 

It remains to be demonstrated, however, whether a single FPGA MD engine can be sufficiently competitive to make it worth developing such a cluster. And if so, how should it be implemented? One thing that is certain is that previous CPU-centric approaches are not viable: long timescales require ultra-short iteration times which make the cost of CPU-device data transfers prohibitive. This leads to another question: is it possible to build such an FPGA MD engine where there is little interaction with other devices?


One advantage we have with current FPGAs is that it is now possible--for simulations of great interest (up to roughly 40K particles)--for all data to reside entirely on-chip for the entire computation. Although this does not necessarily impact performance (double-buffering off-chip transfers still works), it simplifies the implementation and illuminates a fundamental research question: what is the best mapping among particles, cells, and force computation pipelines?  Whereas the previous generation of FPGA/MD systems only dealt with a few cells and pipelines at a time, the concern now is with hundreds of each. Not only does this leads to a new version of the problem of computing pairwise forces with cutoff, explored, e.g., in \cite{Snir04,Shaw05}, it also requires orchestrating RL with the other force computations, and then all of those with motion update and particle movement.

The major contribution is a working end-to-end MD system implemented on a widely used FPGA board.
We have validated simulation quality using Amber 18. In preliminary experiments with the Dihydrofolate Reductase (DFHR) dataset (23.5K particles), the system achieves a throughput of 630ns/day. There are a number of technical contributions.
\begin{itemize}
\item   The first implementation of full MD (RL, LR, and Bonded force with Motion Integration) on a single FPGA, that completely removes the dependency on off-chip devices, thus eliminating the communication overhead of data transfer;
\item   The first analysis of mappings among particles/cells, on-chip memories (BRAMs), on-chip compute units (pipelines) of LR, RL, and bonded forces;
\item   Various microarchitecture contributions related to every aspect of the system, including exploration of RL particle-pair filtering (on-the-fly neighbor lists), two sets of memory architectures (distributed for RL and LR, and global for Bonded), a scoreboarding mechanism that enables motion update in parallel with the force evaluation, and integrating motion update;
\item   Application-aware optimizations through HDL generator scripts.
\end{itemize}
The potential impact is that this system promises to be the basis for long timescale Molecular Dynamics with a commodity cluster.

\section{MD Background}
\label{sec:background}


{\bf Basics.} MD alternates between force calculation and motion update. The forces computed depend on the system being simulated and may include bonded terms, pairwise bond, angle, and dihedral; and non-bonded terms, van der Waals and Coulomb~(e.g., \cite{Haile97}):

\begin{equation}
\mathbf{F^{total}} = F^{bond}+F^{angle}+F^{dihedral}+F^{non-bonded}
\label{eq:LJ_Energy}
\end{equation}

The \textit{Bonded Force} interactions can be expressed as follows: bond (\Cref{eq:bond_force}), angle (\Cref{eq:angle_force,eq:angle_force_appendix}), and dihedral(\Cref{eq:dihedral_force,eq:dihedral_force_appendix}), respectively (from \Cref{eq:LJ_Energy}~\cite{NAMD17c}).

\useshortskip
\begin{equation}
\mathbf{{F}^{bond}_i}= -2k(r_{ij}-r_0)\vec{e_{ij}}
\label{eq:bond_force}
\end{equation}
$\vec{e_{ij}}$ is the unit vector from one item to another, $r_{ij}$ the distance between the two particles, $k$ the spring constant, and $r_0$ the equilibrium distance;

\useshortskip
\begin{equation}
\mathbf{{F}^{angle}_i}= -\frac{2k_{\theta}(\theta-\theta_0)}{r_{ij}}\cdot
\frac{\vec{e_{ij}}cos(\theta)-\vec{e_{kj}}}{sin(\theta)}+f_{ub}
\label{eq:angle_force}
\end{equation}

\useshortskip
\begin{equation}
\mathbf{{f}_{ub}}= -2k_{ub}(r_{ik}-r_{ub})\vec{e_{ik}}
\label{eq:angle_force_appendix}
\end{equation}
$\vec{e_{ij}},\vec{e_{kj}},\vec{e_{ik}}$ are the unit vectors from one item to another, $\theta$ the angle between vectors $\vec{e_{ij}}$ and $\vec{e_{kj}}$, $\theta_0$ the equilibrium angle, $k_{\theta}$ the angle constant, $k_{ub}$ the UreyBradley constant, and $r_{ub}$ the equilibrium distance;

\useshortskip
\begin{equation}
\mathbf{{F}^{dihedral}_i}= -\nabla\frac{U_{d}}{\vec{r}}
\label{eq:dihedral_force}
\end{equation}

\useshortskip
\useshortskip
\begin{equation}
\mathbf{{U}_d}= \left\{
\begin{array}{c}
k(1+cos(n\psi+\phi)) \quad n>0, \\
k(\psi-\phi)^2 \quad\quad n=0.
\end{array}
\right.
\label{eq:dihedral_force_appendix}
\end{equation}
$n$ is the periodicity, $\psi$ the angle between the $(i,j,k)$-plane and the $(j,k,l)$-plane, $\phi$ the phase shift angle, and $k$ the force constant.

The \textit{Non-bonded Force} uses 98\% of FLOPS and includes Lennard-Jones (LJ) and Coulombic terms. For particle $i$, these can be:

\useshortskip
\begin{equation}
\mathbf{{F}^{LJ}_i}= \sum_{j \neq i} \frac{\epsilon_{ab}}{\sigma^2_{ab}}
\left\{ 48 \left( \frac{\sigma_{ab}}{ | r_{ji} | } \right)^{14} - 24 
 \left( \frac{\sigma_{ab}}{ | r_{ji} | } \right)^{8} 
 \right\} \vec{\mathbf{r_{ji}}}
\label{eq:LJ_Force}
\end{equation}

\useshortskip
\begin{equation}
\mathbf{{F}^{C}_i}= \frac{q_{i}}{4\pi}\sum_{j \neq i} \frac{1}{\epsilon_{ab}}\left\{\frac{1}{|r_{ji}|}\right\}^{3}
\vec{\mathbf{r_{ji}}}
\label{eq:C_Force}
\end{equation}
where the $\epsilon_{ab}$ (unit: $kJ$ or $kcal$) and $\sigma_{ab}$ (unit: meter) are parameters related to the types of particles.

The LJ term decays quickly with distance, thus a \textit{cutoff radius, $r_c$,} is applied: the LJ force is zero beyond it. The Coulombic term does not decaying as fast; but this term can be divided into two parts, fast decaying within $r_c$ and slowly developing beyond it. Consequently, we approximate the LJ force and the fast decaying part of the Coulombic force as the~\textit{Range-Limited (RL) force}, and the other part of the Coulombic force force as the~\textit{Long-Range (LR) force}. RL is the more computationally intensive (90\% of flops) and is calculated as:

\useshortskip
\begin{equation}
\frac{\mathbf{F}^{RL}_{ji}}{\mathbf{r_{ji}}} = A_{ab} r_{ji}^{-14} + B_{ab} r_{ji}^{-8} + QQ_{ab} r_{ji}^{-3}
\label{eq:RL_Force}
\end{equation}
where $A_{ab}$ = 48$\epsilon_{ab}$$\sigma_{ab}^{12}$, $B_{ab}$ = -24$\epsilon_{ab}$$\sigma_{ab}^{6}$, $QQ_{ab}$ = $\frac{{q}_{a} {q}_{b}}{4\pi \epsilon_{ab}}$.

The LR force is calculated by solving the Poisson Equation for the given charge distribution.

\useshortskip
\begin{equation}
    \mathbf{F^{LR}_i}=\sum_{j\neq i} \frac{q_j}{|r_{ji}|}
    \vec{\mathbf{r_{ji}}}
    \label{eq:LR_C_Force}
\end{equation}

\begin{equation}
    \rho_g = \sum_p Q_p\phi(|x_g-x_p|)\phi(|y_g-y_p|)\phi(|z_g-z_p|)
\end{equation}

LR is often calculated with a grid-based map of the smoothing function converted from continuous space to a discrete grid coordinate system~\cite{Young09}. Each particle is interpolated to grid points by applying a third-order basis function for charge density calculation. Grid points obtain their charge densities from neighboring particles within a range of two grid points in each direction. There, grid electrostatics are converted into the Fourier domain, evaluated using the Green's function, then converting back through an inverse FFT.


\vspace*{1mm}
\noindent
{\bf Force Evaluation Optimizations.} RL uses the cutoff to reduce the $\mathcal{O}(N^2)$ complexity: forces on each {\it reference particle} are computed only for {\it neighbor particles} within $r_c$. The first approximation is the widely used partitioning of the simulation space into equal sized cells with a size related to $r_c$. The particles can be indexed using \textit{cell-lists}~\cite{Brown11b}: for any reference particle and a cell length of $r_c$, only neighbor particles in the 26 \textit{neighboring cells} need to be evaluated. Another optimization is Newton's 3rd Law (N3L): since the force only needs to be computed once per pair, only a fraction of the neighboring cells need to be referenced.  Most of the particles, however, are still outside the cutoff radius. In CPU implementations this can be handled by periodically creating neighbor lists. In FPGAs, the preferred method is to do this on-the-fly \cite{Chiu09,Chiu10,Chiu11a} through \textit{filtering}.


\vspace*{1mm}
\noindent
{\bf Boundary Conditions.} To constrain particle movement inside a fixed size bounding box, we apply~\textit{Periodic Boundary Conditions (PBC)}. When evaluating particles in boundary cells, we imagine a fictional space that is an exact copy of the simulated space.


\vspace*{1mm}
\noindent
{\bf Motion Integration.}
The change of position and velocity of each particle can be computed using the Verlet algorithm~\cite{Grubmuller91}. Since we are using a short simulation timestep (2 femtoseconds), we can use simple integration equations:
\useshortskip
\begin{equation}
\vec{a}(t) = \frac{\vec{F}(t)}{m}
\label{eq:motion_update_1}
\end{equation}

\useshortskip
\begin{equation}
\vec{v}(t+\Delta t) = \vec{v}(t) + \vec{a}(t)\times\Delta t
\label{eq:motion_update_2}
\end{equation}

\useshortskip
\begin{equation}
\vec{r}(t+\Delta t) = \vec{r}(t) + \vec{v}(t+\Delta t)\times\Delta t
\label{eq:motion_update_3}
\end{equation}
where $m$ is mass, $\vec{a}$ is acceleration, $\vec{v}$ is velocity, $\vec{r}$ is position.



\begin{figure*}[ht]
    \centering
    \includegraphics [width=0.8\textwidth] {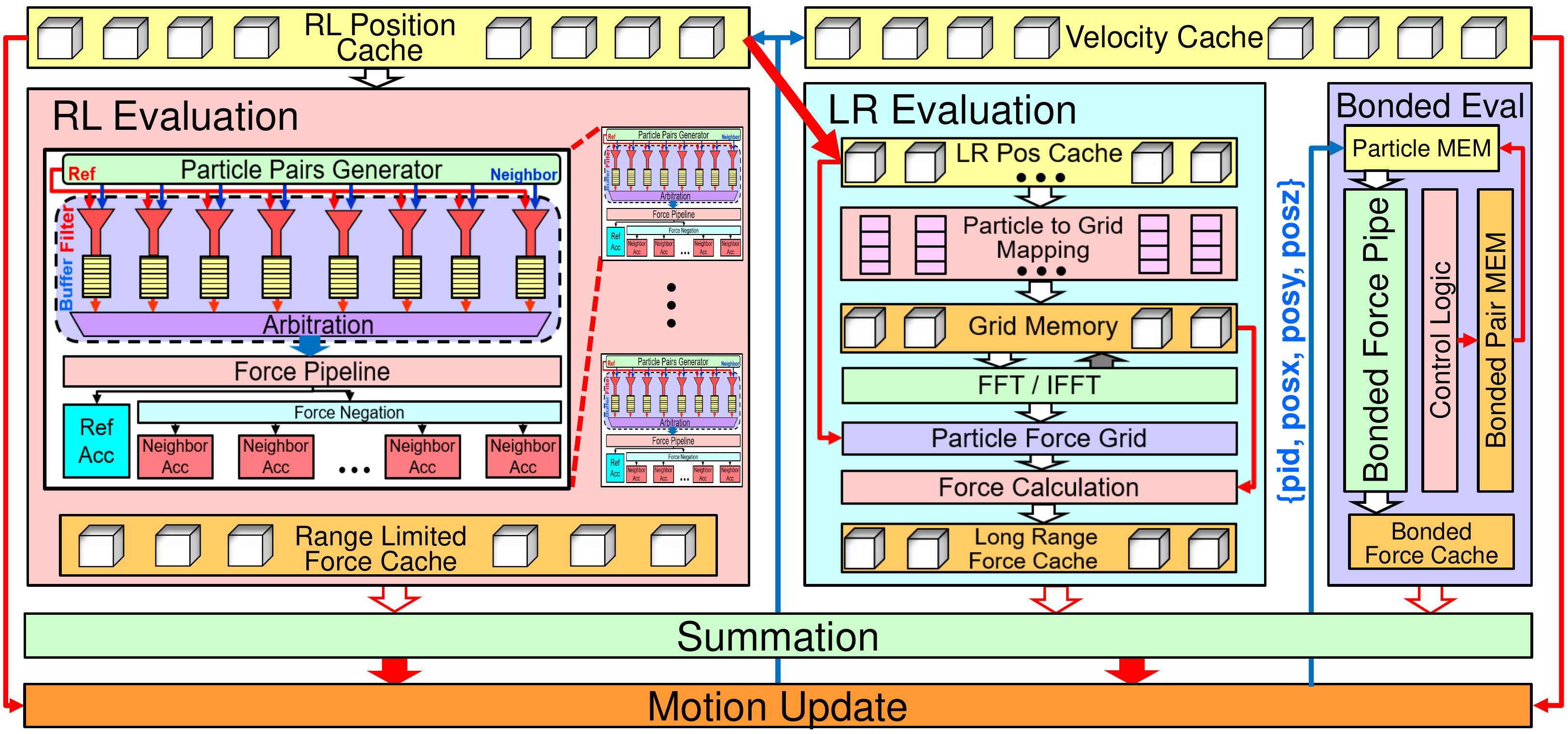}
\vspace*{-0.1truein}
    \caption{MD End-to-End System Overview}
    \label{fig:OverAllArch}
\end{figure*}

\section{FPGA-MD System Architecture}

In this section, we cover the four major components inside an MD simulation system, along with some high-level design decisions. We begin with a classic FPGA-based MD force evaluation pipeline and then add several function units that, in previous implementations, were executed on the host processor or embedded cores.

\subsection{Overall Architecture}

Since configuration time is long with respect to iteration time, the design is largely fixed within a single simulation. A design goal is to give the force computations resources such that their compute times are equalized; resource allocation to summation and motion update is analogous. All components (LR, RL, etc.) have parameterized designs with performance proportional to the parallelism applied (and thus chip resources used). This applies also to fractional parallelism: some components can be \textit{folded}, e.g., to obtain half performance with half resources.

Figure~\ref{fig:OverAllArch} depicts the proposed FPGA-MD system.~\textbf{RL} units evaluate the pair-wise interactions. Since this is the most computationally intensive part, the base module is replicated multiple times.~\textbf{LR} unit includes: (i) mapping particles to a charge grid, (ii) conversion of charge grid to potential grid via 3D FFT (and inverse-FFT), and (iii) evaluating forces on individual particles based on the potential grid. Since our timestep is small (2$fs$), LR is only updated every few iterations.~\textbf{Bonded Evaluation} unit has pipelines for the three parts to evaluate~\Crefrange{eq:bond_force}{eq:dihedral_force_appendix}. At the end of each timestep,~\textbf{Summation} unit sums the three partial forces and sends the result to~\textbf{Motion Update} unit to update position and handle particle migration among adjacent cells. 

\subsection{Particle Memory Choices}

FPGAs provide several modes of data storage, which creates a large design space to be explored.

\vspace*{1mm}
{\bf 1. Which memory resource should we choose?}
The Intel Stratix 10 FPGA~\cite{Intel18b} has 3 types of on-chip memory blocks: eSRAM, M20K, and MLAB. eSRAMs have a large capacity and low latency; but also fixed data width, limited implementation flexibility, and only exists in certain models. MLABs are built on top of the FPGA's logic element, the ALM, which is the resource bottleneck of our implementation and is only suitable for shallow memory arrays.  M20Ks are small SRAM blocks distributed over the entire chip, feature both low latency and a large number of independent ports. The M20Ks are thus the ideal memory for our implementation.

\vspace*{1mm}
{\bf 2. Should a single large memory be used, or multiple memory sets targeting different types of data?}
Each particle has the three sets of information:~\textit{coordinate}, \textit{velocity}, and~\textit{force}, which each have 3 components. Also, storage is needed for the particle type and chemistry indexes. Among the 3 sets of information, their active periods in the datapath are different. Position data are accessed at the beginning, and traverse the entire datapath through force accumulation. Force data is not needed until the force accumulation stage. Velocity data is only needed during motion update. The access pattern is also different. Position data is read-only during the force evaluation phase and gets updated during motion update stage. Force data is actively read and written during force accumulation (since true data dependency cannot be avoided, but can be pipelined, details in Section~\ref{sssec:accumulation}). Velocity data is read once and updated immediately during motion update. Given the above differences, our system implements 3 sets of caches based on the information they carry (position, force, velocity), but they share the same read\&write (R\&W) address for easy management.

\vspace*{1mm}
{\bf 3. Should there be a single large memory module for each of the three types? Or multiple small memory modules?}
As mentioned above, particles are organized into cells. A straightforward solution is to have a large memory with a pre-allocated address space for each cell. This assumes that particles in the dataset are evenly distributed. This design has two advantages: first, the single data source reduces the wiring complexity between the particle cache and force evaluation pipelines; and second, the unified address space simplifies the logic for generating particle pairs for filter evaluation. However, these advantages are not free: a single memory module limits the R\&W bandwidth of the cache, which is then insufficient to feed hundreds of pipelines. A possible solution is to add a small position cache on top of each force pipeline. Another solution is to implement independent memory modules for each cell and each type of data; this solves the bandwidth limitation, but at a cost of complex wiring and slightly worst timing. These schemes will be explored further, along with workload mapping, in the following subsection.

\subsection{RL Architecture}

\begin{figure}[ht]
    \centering
    \includegraphics [width=0.45\textwidth] {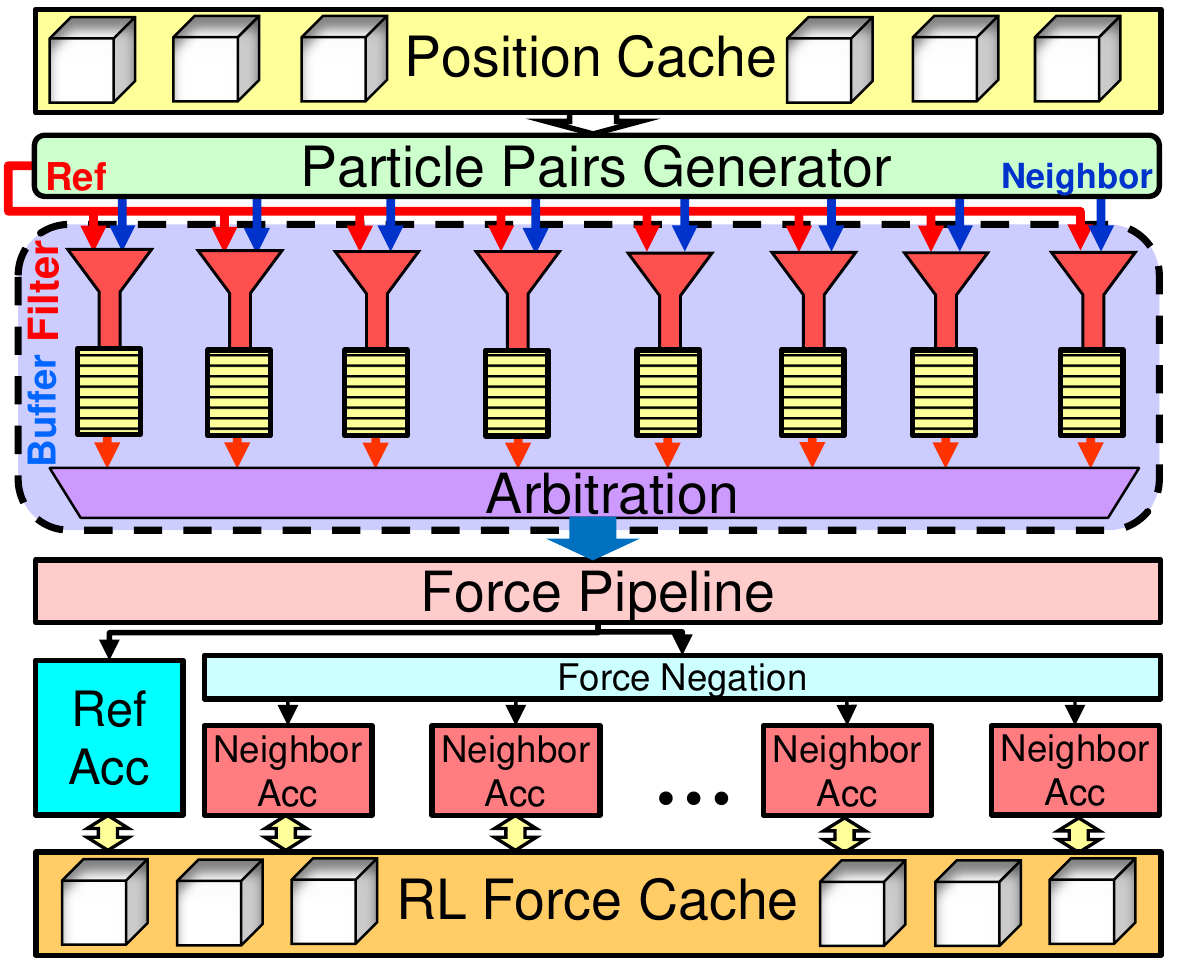}
\vspace*{-0.1truein}
    \caption{RL Evaluation Architecture Overview}
    \label{fig:short_range_pipe}
\end{figure}

In the RL Evaluation pipeline (Figure~\ref{fig:short_range_pipe}) the particle position cache holds the initial position of each particle. Modern high-end FPGAs~\cite{Intel18a} provide enough on-chip storage to hold particle data for our range of simulations. Next is a set of filters that performs a distance evaluation of possible particle pairs (in certain neighboring cells) and only pass the pairs within the cutoff radius, $r_c$. Remaining data then enter the most computationally intensive part in the process: force evaluation. 


\subsubsection{Particle-Pair Filtering}
\label{sssec:RL_filtration}

\begin{figure}[ht]
\centering
    \centering
    \includegraphics[width=1.0\linewidth]{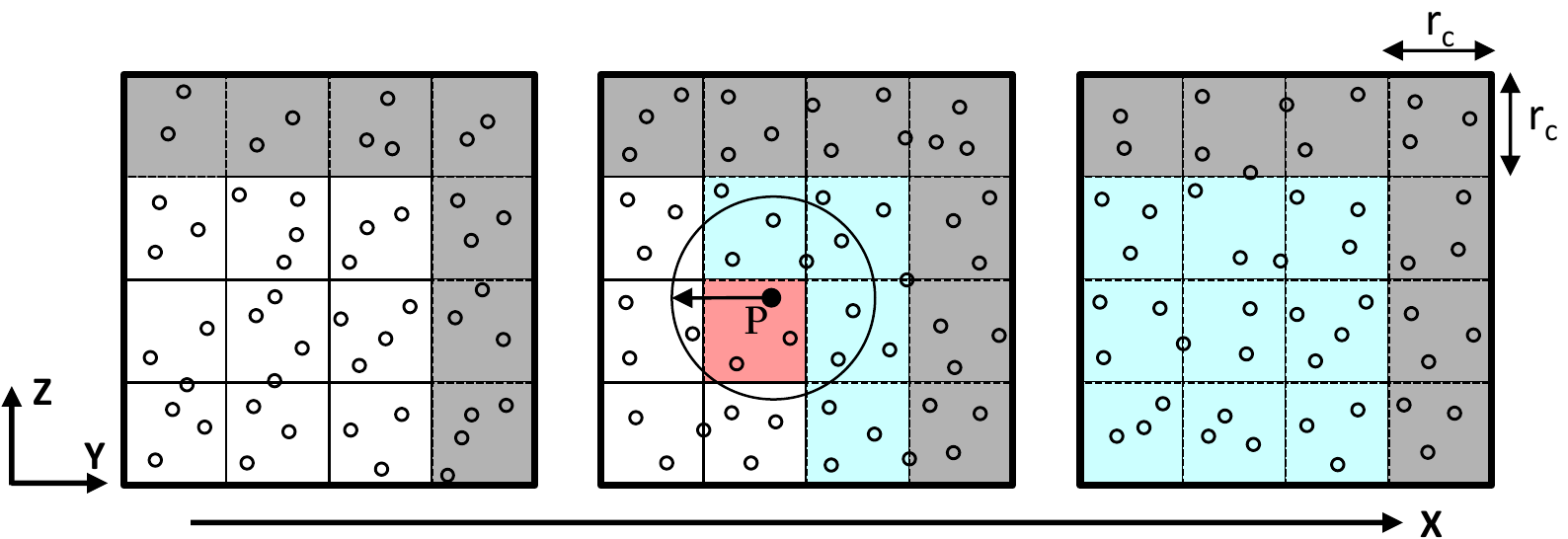}
\vspace*{-0.25truein}
    \captionof{figure}{Simulation space about particle $P$. Its {\it cell neighborhood} is shown in non-gray; cell edge size is the cutoff radius (circle). After application of N3L, we only need to consider half of the neighborcells (blue) plus the homecell (red).}
    \label{fig:celllist}
\end{figure}

Mapping among cells, BRAMs, and filters is complex and is described below. Once a particle pair is generated and sent to a filter, its distance is compared with r$_c$ (actually $r^2$ with $r_c^2$ to avoid the square root). Possible neighbor particles can reside in 27 cells in 3-dimensions (13+1 if considering N3L, as shown in Figure~\ref{fig:celllist}). The average pass rate is not high:
\begin{equation}
    Average\_Pass\_Rate = \frac{\frac{4}{3}\pi r_{c}^{3}}{27\times r_{c}^{3}} = 15.5\%
    \label{eq:filter_passrate}
\end{equation}
Therefore providing a force pipeline with at least one valid output per cycle requires a bank of at least seven filters plus load balancing. We use eight filters per force pipeline. If there are multiple valid outputs, round-robin arbitration is used. The not-selected valid outputs are stored in the filter buffer as shown in Figure~\ref{fig:OverAllArch}.

\subsubsection{Force Evaluation}
\label{ssec:force_eval}

Various trade-offs have been explored in other FPGA/MD work \cite{Gu06a,Gu08a,Chiu08}. These are two of the most important.

\noindent
{\bf Precision and Datatype:} CPU and GPU systems often use a combination of single-precision, double precision, integer, fixed-point, and floating-point. ASIC-based systems have complete flexibility and use non-standard types and precisions. FPGAs have multiple implementation possibilities. If logic cells alone are used, then ASIC-type designs would be preferred for fixed~\cite{Fukushige96,Komeiji97,Chiu10} or floating-point~\cite{Scrofano06a,Chiu11a}. Modern FPGAs, however, also have many thousands of embedded ASIC blocks, viz. DSP and/or floating-point units. So while the arithmetic design space is still substantial, preferred designs are likely to be quantized by these fixed-sized \textit{hard} blocks. We find that, in contrast with earlier FPGA-MD studies, there is less advantage to using integer and fixed point; rather we primarily use the \textit{native floating-point IP core}. For certain computations where accuracy is critical, we also employ fixed-point arithmetic; this is at the cost of high resource overhead (see Section~\ref{sssec:force_pipe_resource}).



\noindent
{\bf Direct Computation vs. Interpolation with Table-lookup:} The RL force calculation requires computing $r^{-3}$, $r^{-8}$ and $r^{-14}$ terms. Since $r^2$ is already provided from the filter unit, a total of 8 DSP units (pipelined) are needed to get these 3 values (based on the force pipeline proposed in~\cite{Chiu10}). Plus, we need 3 extra DSP units to multiply the 3 indexes, $QQ_{ab}$, $A_{ab}$ and $B_{ab}$, with $r^{-3}$, $r^{-14}$ and $r^{-8}$ respectively.

\begin{figure}[ht]
    \centering
    \includegraphics [width=0.47\textwidth] {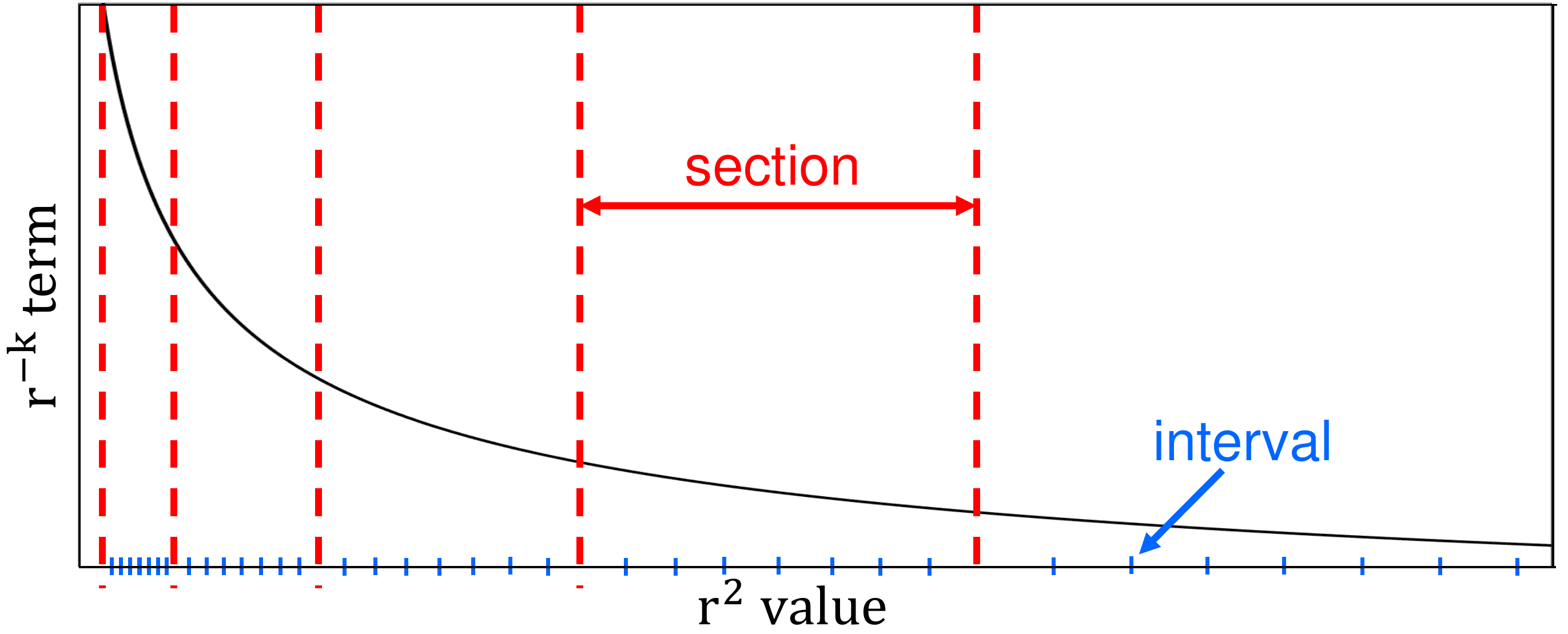}
\vspace*{-0.1truein}
    \caption{Table look-up varies in precision across $r^{-k}$. A fixed number of intervals is assigned to each section.}
    \label{fig:interpolation}
\vspace*{-0.1truein}
\end{figure}

In order to reduce DSP usage, we use interpolation with table-lookup method proposed in~\cite{Gu08a}. As shown in Figure~\ref{fig:interpolation}, we divide the curve into several sections along the X-axis, such that the length of each section is twice that of the previous. Each section has the same number of intervals with equal size. We implement 3 sets of tables for $r^{-3}$, $r^{-8}$ and $r^{-14}$ curve. We use $r^2$, instead of $r$, as the index to further reduce resource consumption that would be needed when evaluating square root and division.

\subsubsection{RL Workload Distribution}
\label{sssec:workload_distribution}

FPGAs provide abundant design flexibility that enables various workload to bare metal mapping schemes. In this subsection, we introduce two levels of mapping: particles onto Block RAMs (BRAMs), and workload onto pipelines. 

\begin{figure}[ht]
    \centering
\vspace*{+0.1truein}
    \includegraphics [width=0.475\textwidth] {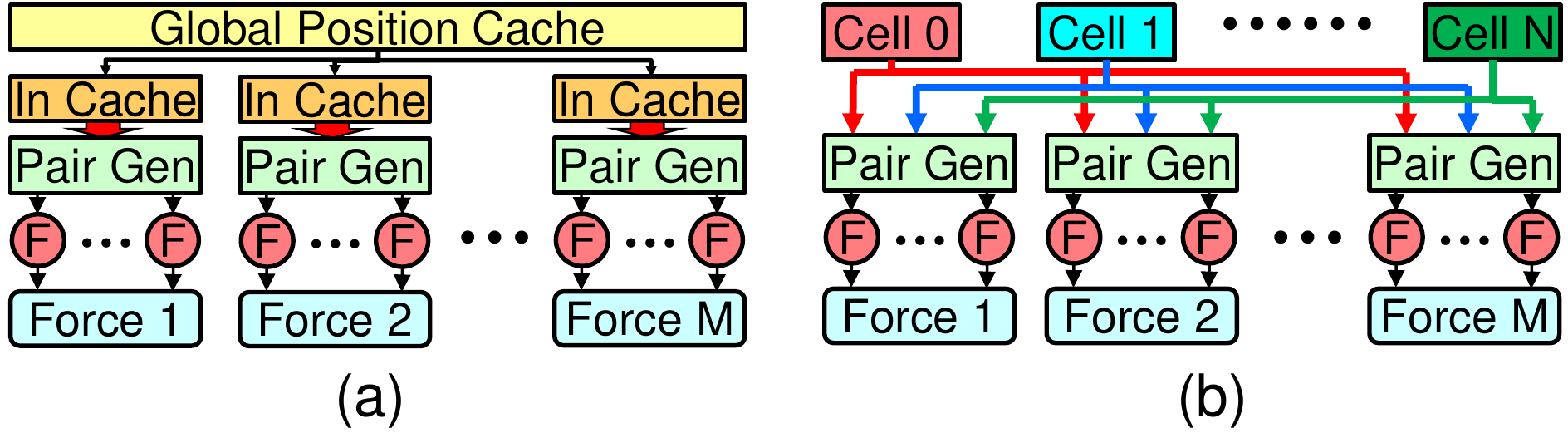}
\vspace*{-0.2truein}
    \caption{Cell to RAM Mapping Schemes: (a) all cells mapped onto a single memory module; (b) each cell occupies an individual memory module.}
    \label{fig:CelltoRAM}
\end{figure}

\noindent
{\bf Cell mapping onto BRAMs}:
Figure~\ref{fig:CelltoRAM} lists two of many possible mapping schemes, which we refer to as~\textit{Mem 1} and~\textit{Mem 2}.

\vspace*{1mm}
\noindent
\begin{itemize}
\item 
\textit{\textbf{Mem 1:}} A single global memory module holds position data for all particles (Figure~\ref{fig:CelltoRAM}a). This design simplifies the wiring between position memory and the hundreds of pipelines. To overcome the bandwidth bottleneck, we insert an input cache at the start of each pipeline to hold the pre-fetched position data. 

\vspace*{1mm}
\noindent
\item 
\textit{\textbf{Mem 2:}} The bandwidth problem can also be overcome by having each cell map onto an individual memory unit (Figure~\ref{fig:CelltoRAM}b). But when there are hundreds of pipelines and cells, the all-to-all connect incurs large resource consumption and timing challenges.
\end{itemize}


\vspace*{1mm}
\noindent
{\bf Workload mapping onto pipelines}: The simulation space is partitioned into cells.  We successively treat each particle in the homecell as a \textit{reference} particle and evaluate the distance with \textit{neighbor} particles from its homecell and 13 neighborcells (N3L). The system then moves to the next cell and so on until the simulation space has been traversed. There are a vast number of potential mapping schemes; due to limited space, we present just three of the most promising.

\begin{figure}[ht]
    \centering
\vspace*{+0.05truein}
    \includegraphics [width=0.48\textwidth] {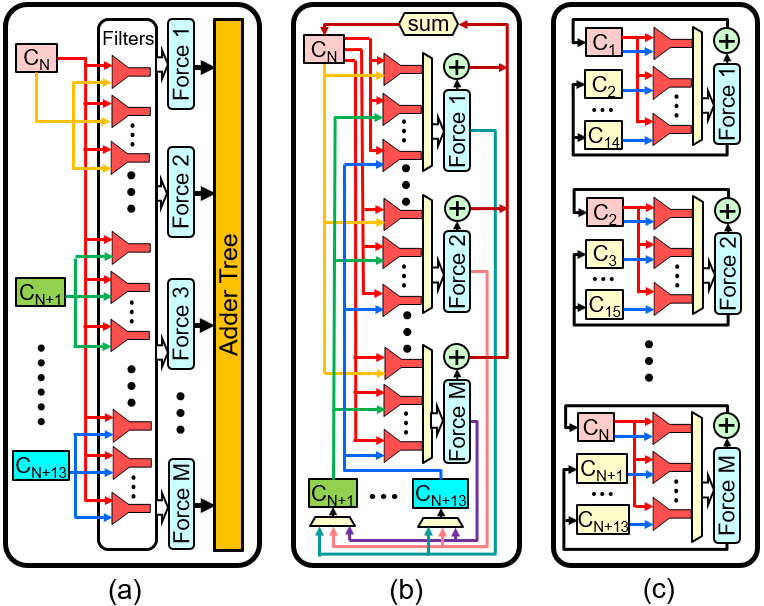}
\vspace*{-0.1truein}
    \caption{Workload mapping onto force pipelines: (a) all pipelines work on the same reference particle; (b) all pipelines work on the same homecell, but with different reference particles; (c) each pipeline works on a different homecell.}
    \label{fig:WorkloadMapping}
\end{figure}

\vspace*{1mm}
\noindent
\begin{itemize}
\item 
\textit{\textbf{Distribution 1: }}\textbf{All pipelines work on the same reference particle (Figure~\ref{fig:WorkloadMapping}a).}
A global controller fetches a particle from the current homecell and broadcasts it to all the filters in the system, around 1000. Potential neighbor particles from home and neighbor cells are evenly distributed among all the filters. The evaluated partial force output from each pipeline is collected by an adder tree for summation and written back. At the same time, the partial forces are also sent back to the neighborcells and accumulated inside each cell. This implementation achieves the workload balance on the particle-pair level. However, it requires extremely high read bandwidth from the position cache to satisfy the need for input data for each filter, and requires high write bandwidth when accumulating partial forces to neighbor particles, since the R\&W only targets 14 cells at a time.

\vspace*{1mm}
\noindent
\item 
\textit{\textbf{Distribution 2: }}\textbf{All pipelines work on the same homecell, but on different reference particles (Figure~\ref{fig:WorkloadMapping}b).} To start, the particle pair generator reads out a reference particle from the homecell for filters belonging to each force pipeline. During the evaluation, the same neighbor particles are broadcast to all filters (belonging to different force pipelines) at the same time, since the neighbor particle set for every reference particle is the same as long as they belong to the same homecell. Compared with the first implementation, this one alleviates the pressure on the read port of the position cache. The tradeoff is that partial forces targeting the same neighbor particle may arrive at the neighborcell at the same time; thus a special unit is needed to handle the read-after-write data dependency. Since each force pipeline is working on different reference particles, an accumulator is needed for each force pipeline.

\vspace*{1mm}
\noindent
\item 
\textit{\textbf{Distribution 3: }}\textbf{Each pipeline works on its own homecell (Figure~\ref{fig:WorkloadMapping}c).} Under this mapping scheme, each filter only needs to interact with a subset of spatially adjacent homecells, along with a set of neighborcells. Compared with the previous two schemes, there is only interaction among a small set of cells. This method not only fully utilizes the parallelism in force evaluation, but also reduces the number of wires between particle caches and force evaluation units. The downside, however, is load balancing. Suppose we have 100 pipelines, but 150 cells. After each pipeline evaluates a cell, half of the pipelines will remain idle while the others evaluate a second homecell. To avoid this waste of resources, an application-aware mapping scheme is required. But this can be overcome by application-aware mapping, which will be covered in Section~\ref{ssec:workload_adoption}.

\end{itemize}

\subsection{LR Architecture}

\begin{figure}[ht]
\vspace*{+0.05truein}
    \centering
    \includegraphics [width=0.4\textwidth] {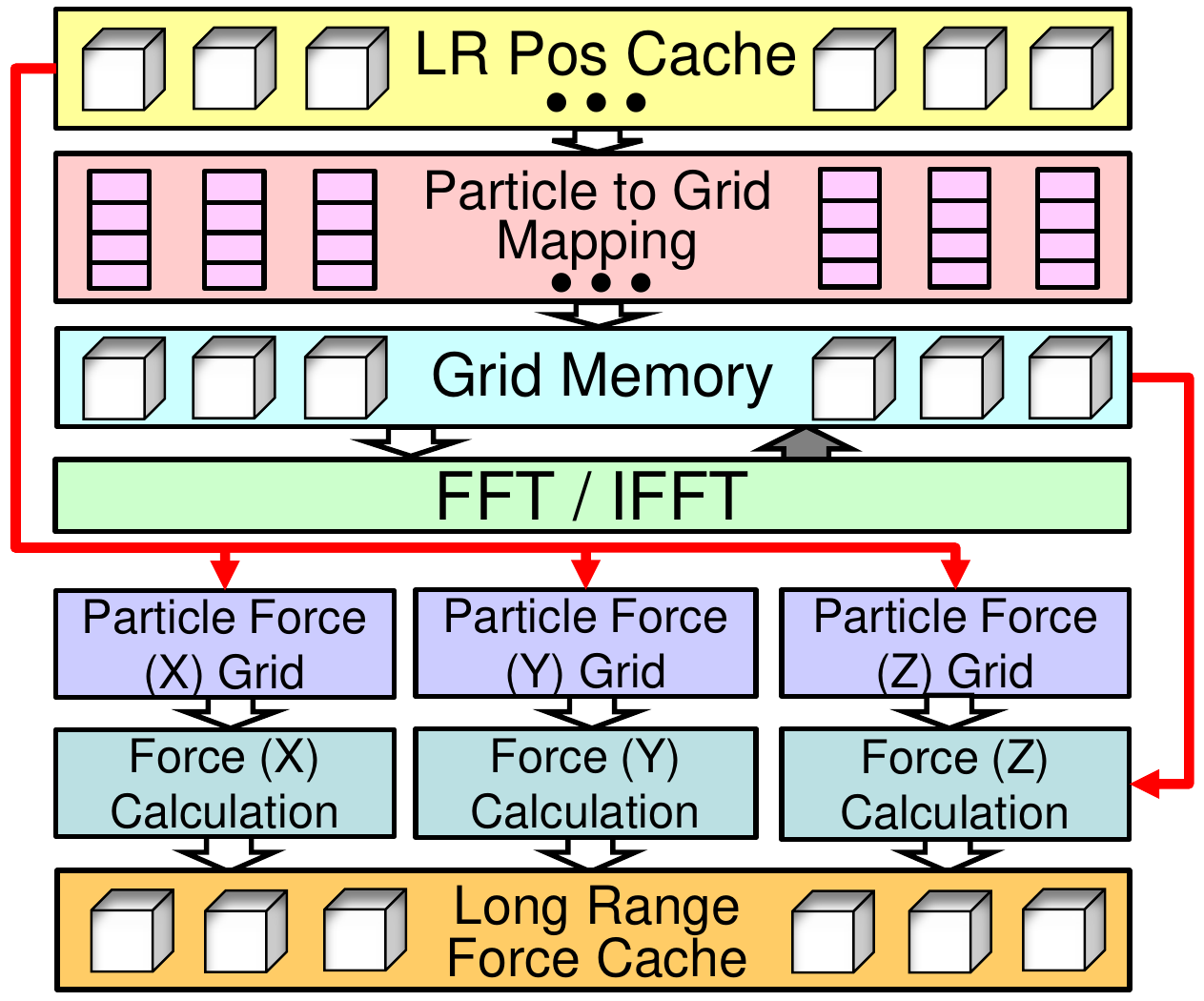}
\vspace*{-0.1truein}
    \caption{LR Evaluation Architecture Overview}
    \label{fig:LRArch}
\end{figure}

LR computation (Figure~\ref{fig:LRArch}) begins with a cache of position data, which maintains particle information when mapping to the particle grid and the force calculation. The position cache is necessary since position data may change during LR evaluation. Particle charges are evaluated and assigned to 64 neighboring cell locations using a third order basis function, with results stored in~\textit{grid memory}. After all particle data is consumed, the FFT evaluation runs on the resulting grid (through each axis X, Y, and Z). Resulting data, after multiplying with the Green's function, is replaced in the memory grid only a few cycles after evaluation. This is possible because of the pipeline implementation of the FFT. The inverse FFT is then performed on each dimension. Finally, forces are calculated for each individual particle using the final FFT grid results and the starting particle position information saved previously in the position cache. These are then saved into a force cache which is used during the motion update phase to apply long-range forces to the particle positions. 

\begin{figure}[ht]
    \centering
    \includegraphics [width=0.45\textwidth] {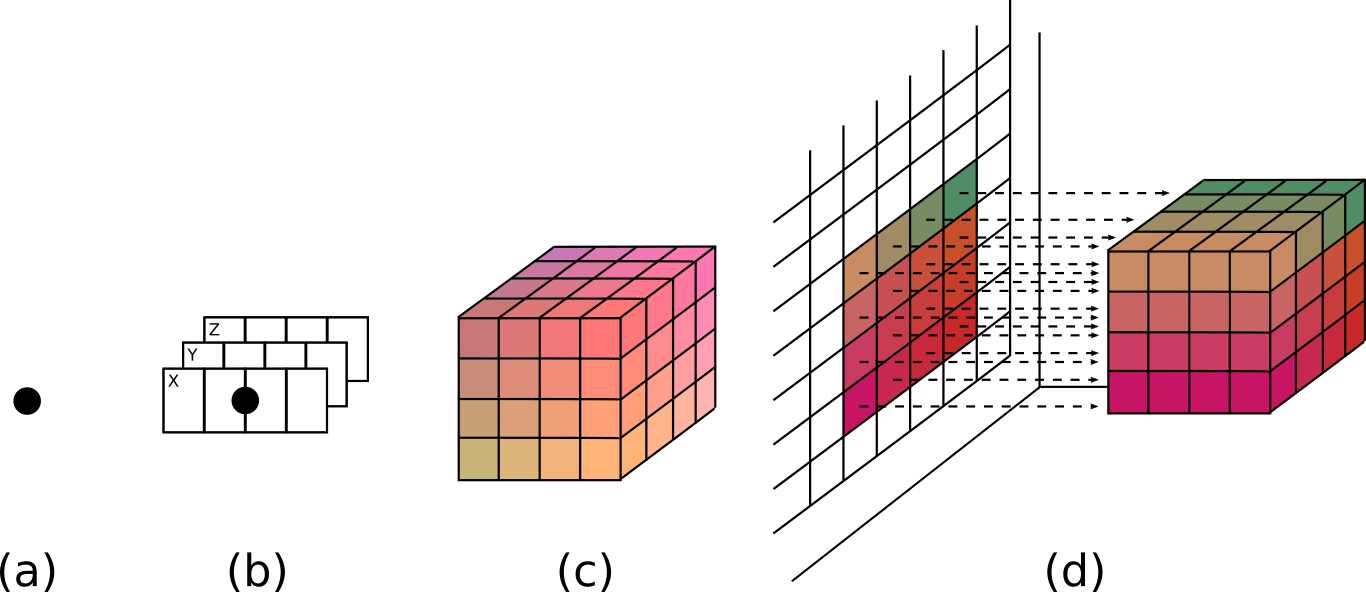}
\vspace*{-0.1truein}
    \caption{Particle to grid flow: (a) Initial particle position data; (b) Particle to 1D interpolation for each dimension using basis functions; (c) Mapping 1D interpolation results to a 4x4x4 3D grid; (d) Final 64 grid points to 16 independent memory banks}
    \label{fig:Particle_Grid_64}
\end{figure}

\subsubsection{Particle to Grid Mapping}
\label{Particle_To_Grid_Mapping}
The third order basis functions \Cref{eq:basis function} are used to spread particle charges to the four closest grid points, based on particle position data, and can be independently evaluated for each dimension \cite{Gu07a}. After a particle is evaluated in each dimension, values are assigned to 64 neighboring cells and each result is accumulated into grid memory locations. Figure \ref{fig:Particle_Grid_64} shows the process of a single particle's influence on 64 neighborcells and their mapping to the grid memory structure. Parallel particle-to-grid mapping occurs with the use of accumulators before entering grid memory due to restrictions in using BRAMs \cite{VanCourt06d,Sanaullah16}. 

\begin{equation}
\left\{
\begin{aligned}
    \phi_0(oi) &= -1/2oi^3 + oi^2 - 1/2oi\\
    \phi_1(oi) &= 3/2oi^3 + 5/2oi^2 + 1\\
    \phi_2(oi) &= -3/2oi^3 + 2oi^2 + 1/2oi\\
    \phi_3(oi) &= 1/2oi^3 - oi^2.
    \end{aligned}
    \right.
    \label{eq:basis function}
\end{equation}

\subsubsection{Grid Memory}
\label{Grid_Memory}
We store grid points in BRAMs using an interleaved memory structure.
This allows for stall-free access of grid locations while performing FFT calculations.

\subsubsection{FFT}
\label{FFT}
The FFT subsystem performs calculations in parallel using vendor supplied FFT cores. The FFT units are assigned to specific banks of the grid memory to ensure high throughput memory access. As a result, grid data can be continuously streamed through all FFT cores in parallel. While output is being generated for a given vector, a new input is sent for the next set of calculations. Each dimension is performed sequentially until all three dimensions are completed on the memory grid. Once all three dimensions are evaluated and converted into the Fourier-domain, the grid is multiplied with Green's function, before proceeding to the inverse FFT stage going through each dimension again and converting back. Final values at each grid point are used to compute the LR force for each particle based on its position.

\subsection{Bonded Architecture}
\label{ssec:bonded_arch}

\begin{figure}[ht]
    \centering
    \includegraphics [width=0.47\textwidth] {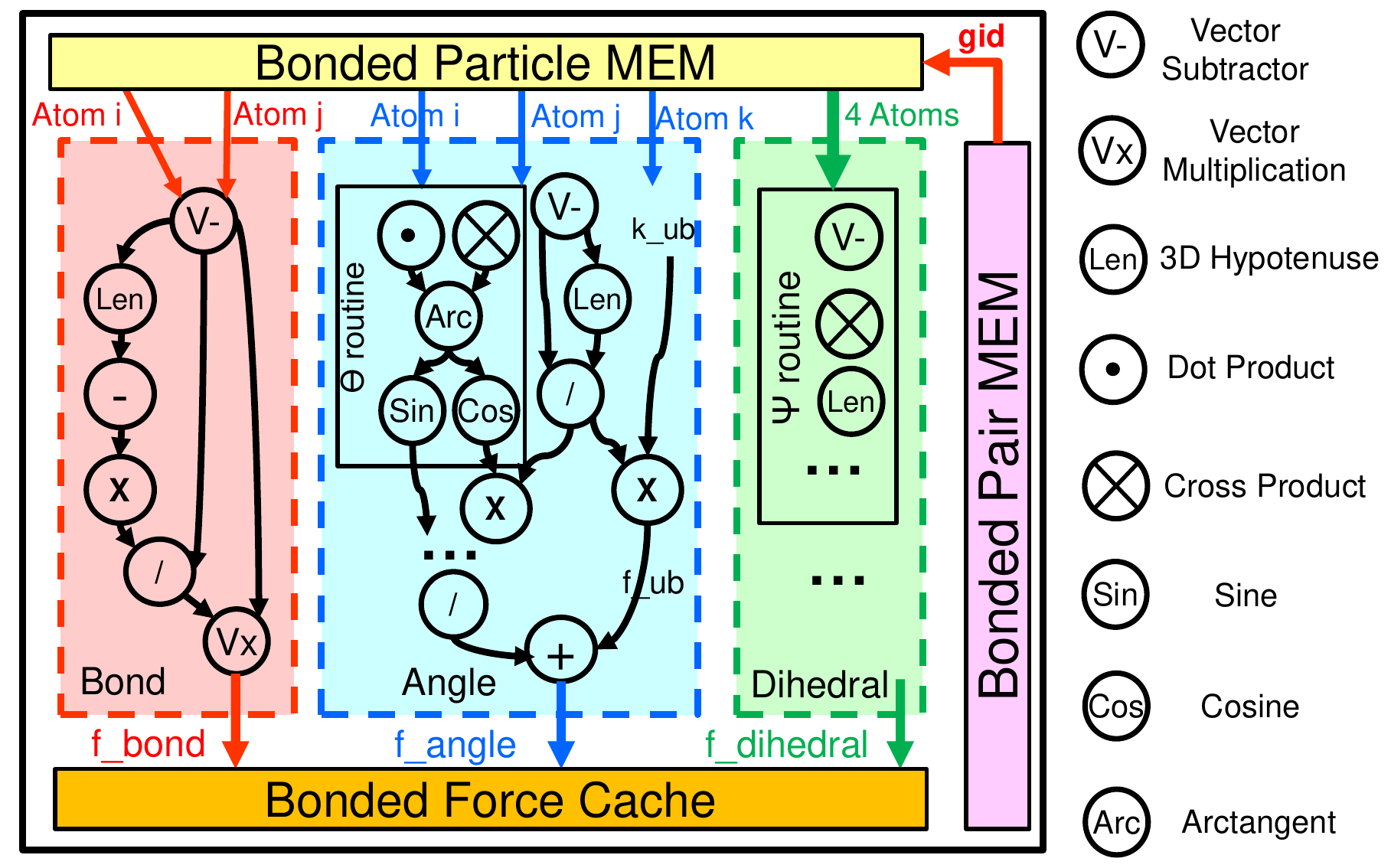}
\vspace*{-0.1truein}
    \caption{Bonded Force Evaluation Architecture}
    \label{fig:bonded_arch}
\end{figure}

\subsubsection{Sequential Evaluation of Bonded Interactions}

As shown in Figure~\ref{fig:bonded_arch}, we evaluate three types of bonded interactions: bond, angle, and dihedral, which have, respectively contributions from 2, 3, and 4 atoms. For a given dataset, the covalent bonds remain fixed as long as no chemical reaction is involved. In general, the bonded computation requires only a few percent of the FLOPs, so attenuation rather than parallelism is advantageous: we therefore process bonds sequentially.


\subsubsection{Bonded Force Memory Architecture}

For LR and RL we organize the particle data based on cells; this proves costly for bonded force evaluation.  Rather than particles interacting with others based on their spatial locality, bonded interactions have a fixed set of contributing particles. As simulation progresses, particles can move across different cells and require extra logic to keep track of their latest memory address. Given the fact that we process bondeds sequentially, and that this requires little memory bandwidth, we propose a different memory architecture: a single global memory module (\textit{Bonded Particle MEM} in Figure~\ref{fig:bonded_arch}) that maintains information on each particle position based on a fixed particle global id (gid). The gid is assigned prior to the simulation and remains fixed. 

A read-only memory, \textit{Bonded Pair MEM}, holds pairs of gids that form chemical bonds in the dataset. During force evaluation, the controller first fetches a pair of gids along with other parameters from Pair MEM, then proceeds to fetch the related particle position from Particle MEM and sends this for force evaluation. The evaluated bonded force is accumulated in the~\textit{Bonded Force Cache} addressed by~\textit{gid}. During motion update, the accumulated bonded force summed with partial results from RL and LR. Finally, Particle MEM receives the updated position, along with the particle gid, to maintain an up-to-date value.

\subsection{Force Summation and Motion Integration}

The three force components must be combined before the motion update. Even with load balancing, RL always finishes last; this is guaranteed, in part, by the small variance in the other computations. Therefore we can assume that the LR and bonded force caches always have data ready. Thus, as soon as RL of a certain particle is ready, we can perform the summation and motion update. As described in Section~\ref{sssec:RL_filtration}, for any given particle, it needs to be evaluated with respect to each neighbor particle from 27 cells. Since we make use of N3L to avoid revisiting particle pairs more than once, we need to keep track of how many times each cell has been visited (as homecell and neighborcells). To handle this we propose a \textit{Score Boarding} mechanism. Once computations on all particles in a cell have finished, the Score Board module will access LR, RL and Bounded forces from the corresponding caches for force summation. By doing so, the positions of particles from the same cell can be updated immediately when a cell is fully evaluated; the motion update is executed in parallel with force evaluation with limited resource overhead; a large fraction of motion update latency can therefore be hidden.

After summation for a particle is finished, the aggregated force is sent to the motion update unit, along with particle's position and velocity. Since we organize particle data based on the cells they belong to (except for the bonded unit), particles can move from one cell to another. This creates a challenge on particle memory management: we need to maintain a record of which memory is ready for receiving new particles (due to particles left in the current cell, or the pre-allocated vacant memory space in each cell).  It may take multiple cycles to find an available memory slot when the cell memory is almost full, or to find a valid particle in the cell when the cell memory is almost empty. Our solution is to double buffering the particle position and velocity caches; details are given in Section~\ref{sssec:double_buffer_imple}.


\section{MD System Implementation}
\label{sec:impl}

In this section, we highlight a selection of implementation details.

\subsection{Datatype and Particle Cache}

The system maintains three sets of information for each particle: position, velocity, and force. The first two need to be maintained throughout the entire simulation, while the force data is flushed after motion update.

\subsubsection{RL Particle Cache}
\label{sssec:double_buffer_imple}


\begin{figure}[ht]
\centering
    \centering
\vspace*{-0.1truein}
    \includegraphics[width=0.4\textwidth]{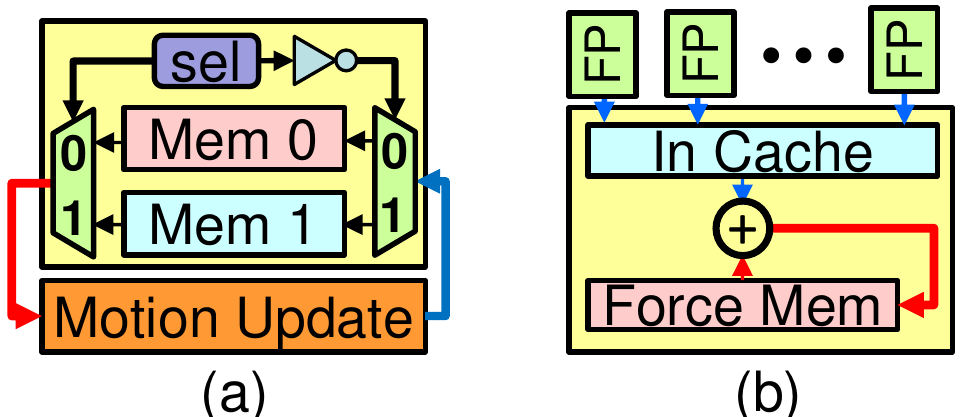}
\vspace*{-0.1truein}
    \captionof{figure}{(a) Double buffer mechanism inside position and velocity cache; (b) Force cache with accumulator.}
    \label{fig:particle_cache}
\end{figure}


\noindent
\textbf{RL Position Cache} organizes data into cells. Double buffering is implemented (Figure~\ref{fig:particle_cache}a)  with the particle \textit{gid}s being kept along with position data, which is used during summation and motion update process.
\textbf{RL Force Cache} is read and written during force evaluation. Since the system has hundreds of pipelines, that many partial forces must be accumulated each cycle. To manage the potential data hazards, we implement an accumulator inside each force cache module (see Figure~\ref{fig:particle_cache}b). After the aggregated forces are read out during motion update, they are cleared for the next iteration. 

\subsubsection{LR Particle Cache}
The LR force evaluation is generally performed every two to four iterations while motion update happens every iteration. Since LR evaluation needs the positions to remain fixed, we allocate a separate LR particle cache (Figure~\ref{fig:OverAllArch} \&~\ref{fig:LRArch}). Every time LR starts a new evaluation (every second iteration in our experiments), it first performs a memory copy from RL Position Cache. To shorten the memory copy latency, we implement LR cache using~\textit{Mem 2}, which provides high write bandwidth.


\subsection{RL Force Evaluation}

\subsubsection{Tree-based One-to-all Mux}
\label{sssec:mux_tree}

As discussed in Section~\ref{sssec:workload_distribution} and shown in Figure~\ref{fig:WorkloadMapping}, when~\textit{Distribution 1 or 2} is selected, each pipeline takes input data from every one of the memory units at a certain stage of evaluation. Essentially a $ M \times P$ switch is needed, where $M$ is the number of memory units and $P$ is the number of pipelines in the system. The large fan-out on memory outputs has a bad impact on operating frequency. Our timing analysis shows that under the direct connection, the system can only run at around 100 MHz, while the force pipeline has a performance of 350 MHz. 

To fill the gap between the two units, we propose a tree-based reduction unit. Each large multiplexer is composed of small registered 1-to-5, 1-to-4 muxes. For example, when module broadcasts to 100 receivers, the 1-to-100 mux is composed of 3 levels of small registers: the first has a single 1-to-4 mux, the second has four 1-to-5 muxes, and the last stage has 20 1-to-5 muxes. In this way, the new mux has a 3 cycle latency but runs at a much higher frequency. Since the datapath is fully pipelined, running at a higher frequency clearly brings more benefit. For~\textit{Distribution 1}, only one set of the mux tree is needed, while~\textit{Distribution 2} requires $M$ copies; the latter limits the number of force pipelines we can instantiate.

\subsubsection{Filter Logic}
\label{ssec:filter_implemetation}

We propose two methods.

\vspace*{1mm}
\noindent
{\bf 1. Filter v1: Direct computation} uses 8 DSP units to calculate $r^2$ in floating-point and compare with $r_c^2$. If the distance is within cutoff, the evaluated $r^2$ is reused by the force pipeline. Since there are 8 filters per pipeline, the direct implementation consumes 48 DSP units, which limits the number of pipelines per chip. 

\vspace*{1mm}
\noindent
{\bf 2. Filter v2: Planar method} uses~\Crefrange{eq:acceleration1}{eq:acceleration3}; note that the $r_c$ terms are constants and not computed. 
\useshortskip
\begin{equation}
    |x|<r_c,|y|<r_c,|z|<r_c
    \label{eq:acceleration1}
\end{equation}
\useshortskip
\begin{equation}
    |x|+|y|<\sqrt{2}r_c,|x|+|z|<\sqrt{2}r_c,|y|+|z|<\sqrt{2}r_c
    \label{eq:acceleration2}
\end{equation}
\useshortskip
\begin{equation}
    |x|+|y|+|z|<\sqrt{3}r_c
    \label{eq:acceleration3}
\end{equation}
To avoid using DSPs altogether, input data is converted from floating-point to 28-bit fixed-point.

\subsubsection{Filter Arbitration}

Round-robin is used to select among filters with a valid output. In order to reduce the latency in the filter bank, which also saves buffer space, the following algorithm delivers one result per cycle.

\vspace*{2mm}
\noindent
1: Shift the current arbitration result left 1 bit, then subtract 1\\
2: Perform NOT on Step 1\\
3: Get valid mask based on data availability in each filter buffer\\
4: Perform AND on Step 2 and Step 3\\
5: Perform 2s compliment on Step 4\\
6: Perform AND on Step 4 \& 5, this is \textbf{new arbitration result}\\
7: If current valid mask only has MSB as 1, then omit Step 1\\
8: If current arbitration result is 0, skip Steps 1-4

\vspace*{2mm}

\subsubsection{RL Force Pipeline}

\begin{figure}[ht]
\centering
    \centering
    \includegraphics[width=0.45\textwidth]{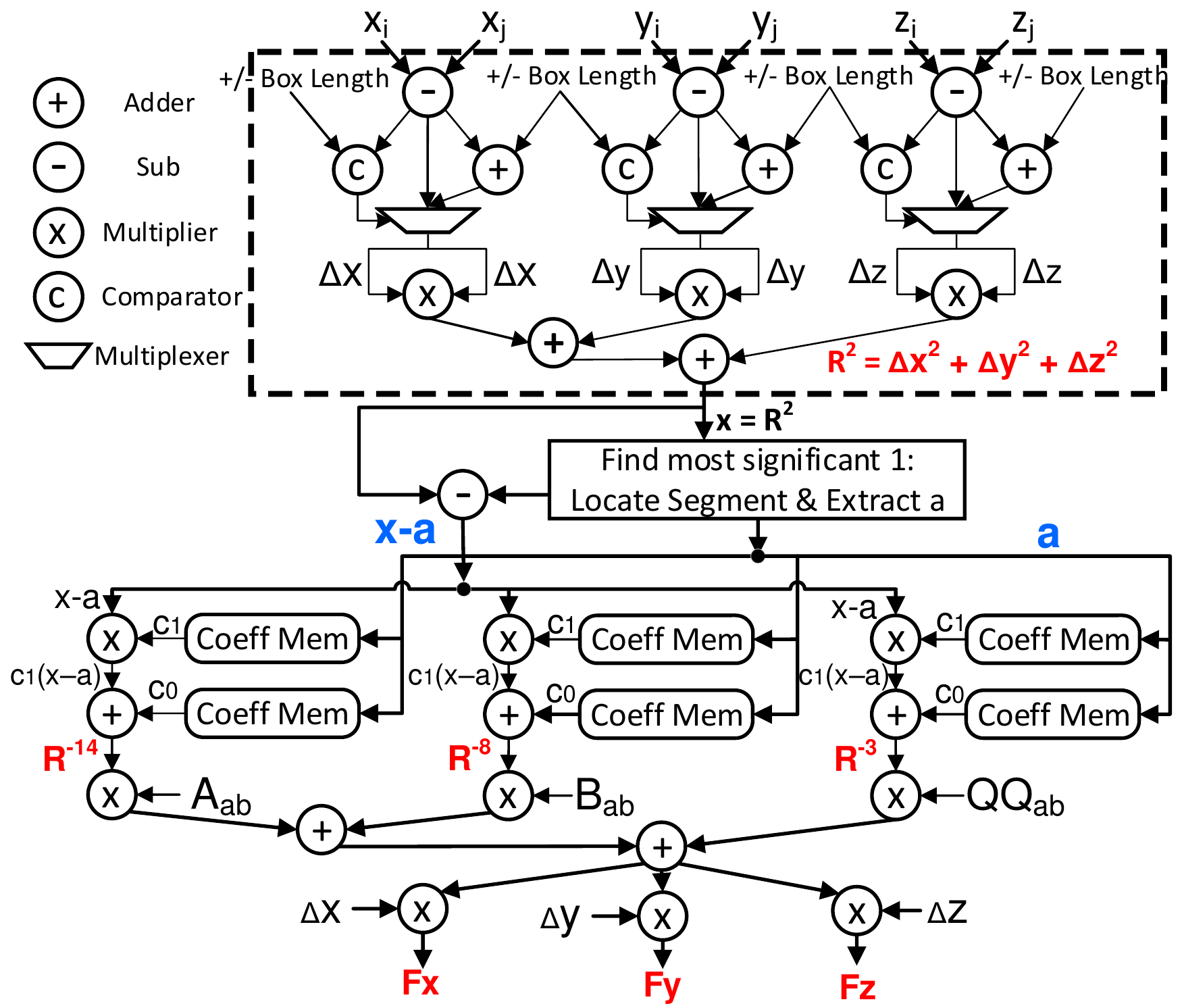}
\vspace*{-0.1truein}
    \captionof{figure}{Force evaluation with first order interpolation.}
    \label{fig:force_pipe}
\end{figure}

Depending on the filter implementation, the force pipeline will receive one of two different inputs. Filter v1 provides $r^2$, while Filter v2 provides only the raw particle position so $r^2$ must be computed (Figure~\ref{fig:force_pipe}). 

The pipeline evaluates forces via interpolation with table-lookup. Assuming the interpolation is second order, it has the format:

\useshortskip
\begin{equation}
    r^{k} = (C_2(x-a)+C_1)(x-a)+C_0
    \label{eq:interpolation}
\end{equation}
where $x=r^2$, $a$ is the $x$ value at the beginning of the target interval, and $x-a$ is the offset into the interval. Based on different datasets, the interpolation coefficients are pre-calculated, packed into the mif file, and loaded onto the FPGA along with position and velocity data. 
After the coefficients are read from memory, the pipeline performs the evaluation following ~\Cref{eq:interpolation}. Figure~\ref{fig:force_pipe} shows this for the first order; the actual system supports upto the third order. 

\subsubsection{Partial Force Accumulator}
\label{sssec:accumulation}

The system traverses particles in the simulation space following cell order. When treating a particle as a reference particle, the filter will check the distance with all the potential neighbor particles residing in the home and neighborcells. We use N3L to avoid evaluating the same particle pair twice. But this also means the evaluated force need to accumulate to 2 particles. A difficulty is that the floating-point adder on FPGA has a latency of 3 cycles, leading to the classic accumulator problem: we can only perform one accumulation every 3 cycles. This is clearly unacceptable. We have two solutions, one for reference particles and one for neighbor particles.

\begin{figure}[ht]
\centering
    \centering
    \includegraphics[width=0.45\textwidth]{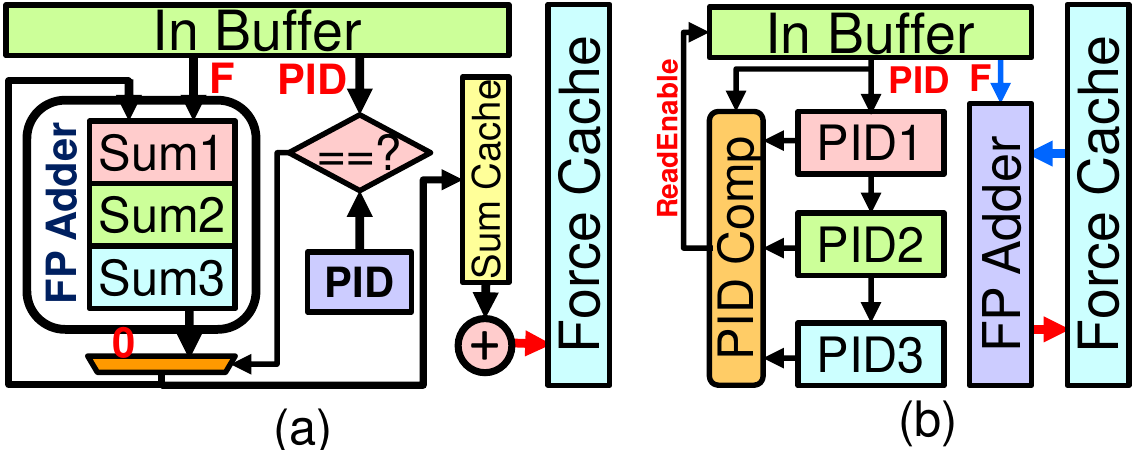}
\vspace*{-0.1truein}
    \captionof{figure}{(a) Accumulator for reference particles, located at the output of the force pipeline; (b) Accumulator for neighbor particles, located at the input of the force cache.}
    \label{fig:accumulator}
\end{figure}

\textbf{Reference Particle Accumulator:}
The pipeline works on the same reference particle until all the neighbor particles have been covered. For reference particles, this requires accumulating to the same value for consecutive cycles. Based on the workload distribution scheme introduced in Section~\ref{sssec:workload_distribution}, the accumulator can be different. \textit{Distribution 1} has multiple pipelines working on the same reference particle and thus needs a tree-based accumulator array to sum the partial result from hundreds of pipelines. For \textit{Distribution 2 or 3} the accumulator must be on the output side of the force pipeline, as shown in Figure~\ref{fig:accumulator}(a). Traditionally, connecting the adder output back to the input forms a simple accumulator. But since the adder has a 3 cycles latency, writing back the output will have 3 temporary sum values cycling inside. 

The solution is as follows: when a valid force arrives, the reference particle id (PID) is checked, if the particle has the same id as the accumulator is working on, then it is forwarded to the adder. If not, then it is recognized as the new reference particle, and the PID register will record the new particle ID. The returned value to the input sets to 0 for 3 cycles to reset the sum value. In the meantime, the output writes to a small cache for 3 cycles, where they are added together and written back to force cache. The design is replicated 3 times to simultaneously handle $F_x, F_y, F_z$. 

\textbf{Neighbor Particle Accumulator:}
Multiple pipelines' outputs may target the same reference particle at the same time. But it is unrealistic to handle those conflicts in each pipeline. One potential solution is to maintain a local copy of the particle result for each neighbor particle and sum those together at the end of an iteration. But this brings large overhead on resources (force caches, additional adder-trees for summing up, etc.) and latency (sum up all particles one-by-one). The solution is to put the neighbor accumulator inside force cache; the pipelines just forward their results to the targeted force cache that holds the neighbor particle. 

The neighbor accumulator is shown in Figure~\ref{fig:accumulator}(b). All incoming data are buffered. Three registers are used to record the currently evaluated particle id. At the beginning of each cycle, a new datum emerges from the buffer and is checked to see if it has the same PID as current active ones. If not, then the partial force is read out from force cache and sent to the pipeline for accumulation. Otherwise, the particle is written back to the bottom of the input buffer (to avoid further congestion in the pipeline). Similar to the reference accumulator, the unit is also replicated three times in each force cache to handle the three force components in parallel.

\subsection{LR Force Evaluation}
\label{ssec:LR_imple}
\subsubsection{Particle to Grid Mapping}
\label{Particle_Grid_Mapping_Impl}
\begin{figure}[ht]
\centering
    \centering
    \includegraphics[width=0.3\textwidth]{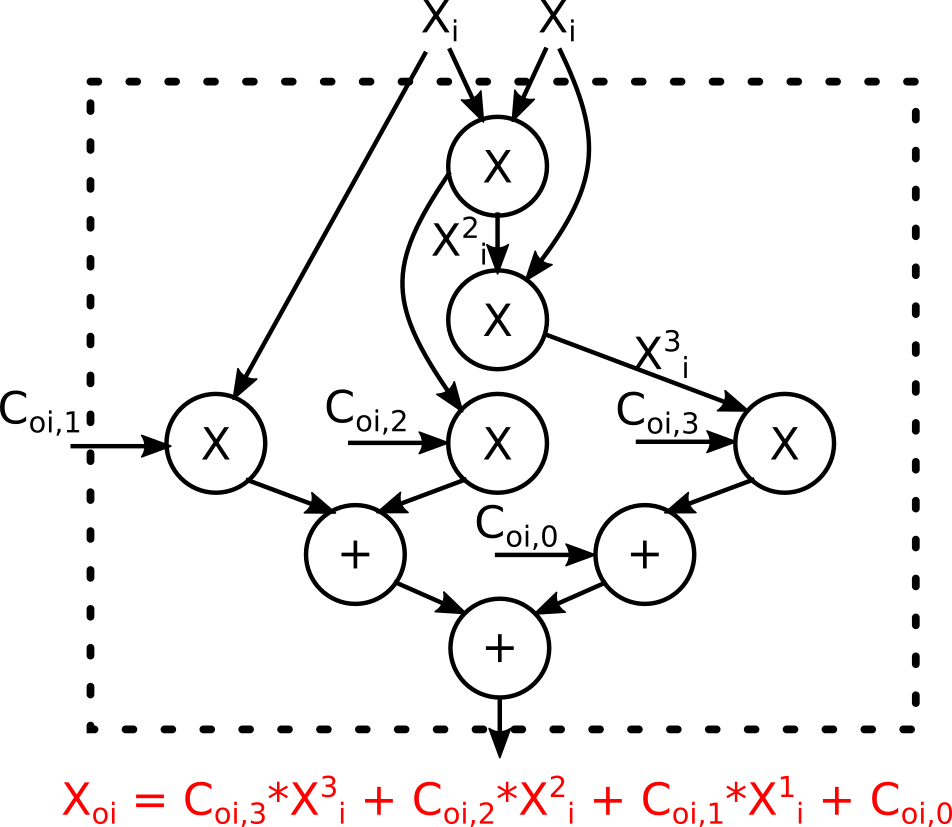}
\vspace*{-0.1truein}
    \captionof{figure}{One instance of the particle to grid conversion equation: The unit is replicated 12 times. Four instances represent the four basis equations for each dimension X, Y, and Z.}
    \label{fig:coeff_impl}
\end{figure}

Due to the large number of particles, the particle to grid mapping must be optimized to avoid adding additional stall cycles when each particle enters the system. This means replication is a must to avoid long delays. The first step is to evaluate each individual basis function per dimension to obtain a single particle contribution to an individual cell. As Figure~\ref{fig:coeff_impl} shows, one function takes 5 steps to evaluate a single equation. This unit can be replicated to evaluate all 4 functions simultaneously and each dimension is done in parallel requiring a total of 12 replications of each unit. After all functions are evaluated, values are combined to form 64 unique values representing the 4x4x4 neighbor grid of cells around the particle. These 64 cells are then accumulated with the previous information, found in their respective cells.

\subsubsection{FFT}
\label{FFT_Impl}

Using the interleaved structure of the grid memory, the FFT implementation allows for the use of multiple FFT units to evaluate each dimension in parallel. 
Since this part of LR is not the bottleneck, a modest number of FFT blocks (16) is currently used.

\subsubsection{Matching RL performance}

\label{Parameterized_LR}
By using a parameterized design, our sample implementation maintains a 2:1 timing ration between LR and RL. Details are complex, but entail using methods such as folding and reusing logic.

\subsection{Bonded Force Pipeline}
\label{ssec:bonded_pipeline}

\begin{figure}[ht]
\centering
    \centering
    \includegraphics[width=0.45\textwidth]{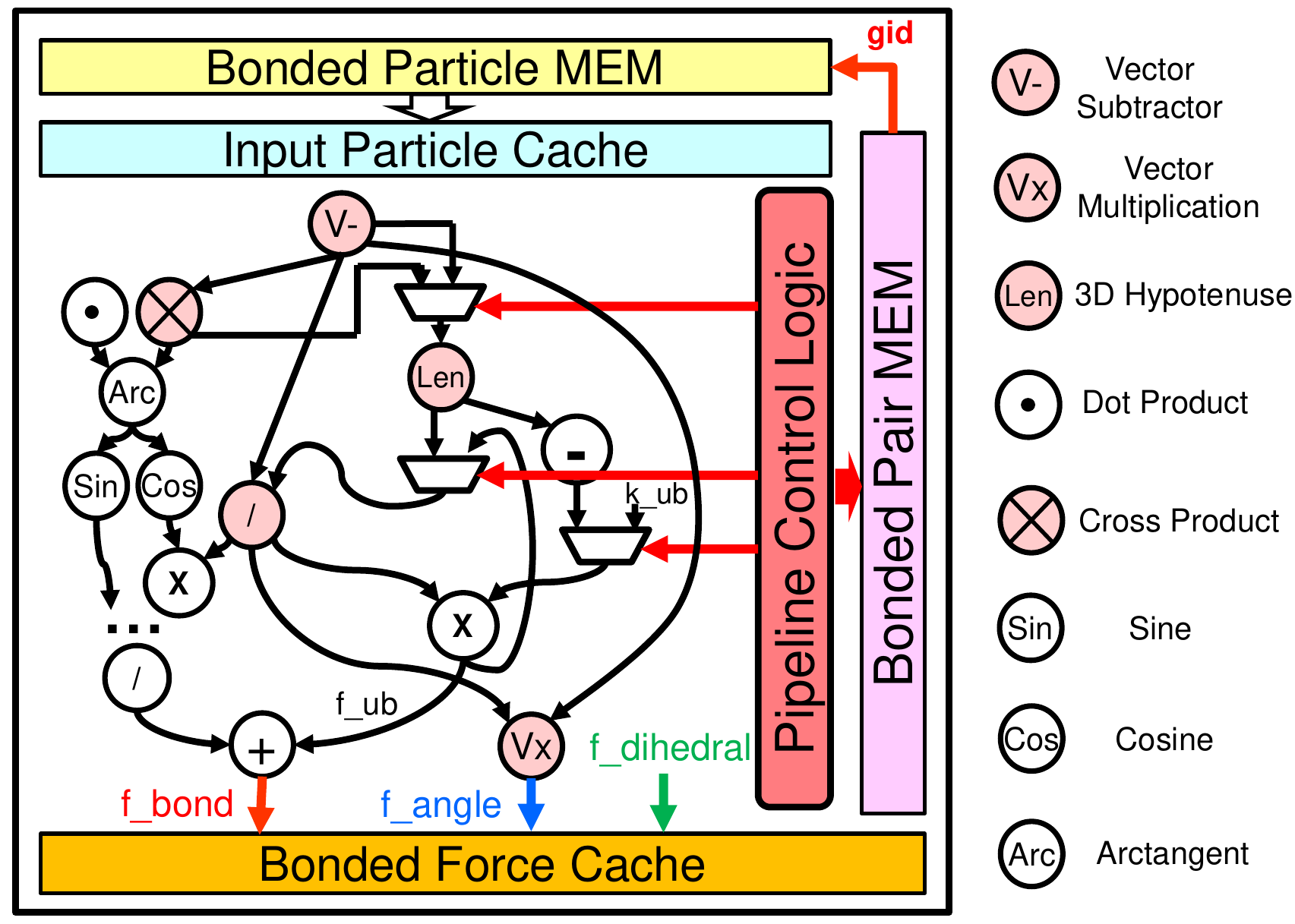}
\vspace*{-0.1truein}
    \captionof{figure}{Motion Update Pipeline}
    \label{fig:bonded_merge_pipe}
\end{figure}

It is possible to stay within the time budget even if only one of the three evaluation pipelines (Figure~\ref{fig:bonded_arch}) is active in a given cycle. Also, many functions overlap among the three interactions. Therefore, to maximize the DSP units' utilization ratio, we merge the three pipelines into a single one with control registers and muxes at different stages of the pipeline (Figure~\ref{fig:bonded_merge_pipe}). 



\subsection{Summation Logic}

\begin{figure}[ht]
\centering
    \centering
    \includegraphics[width=0.4\textwidth]{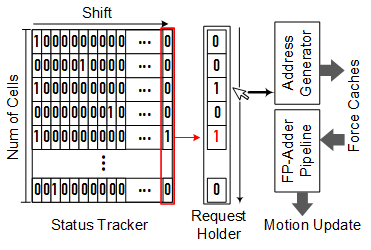}
\vspace*{-0.2truein}
    \captionof{figure}{Summation logic with score boarding mechanism}
    \label{fig:scoreboard}
\end{figure}

\noindent
Summation logic scoreboard support is shown in Figure~\ref{fig:scoreboard}.  The \textit{Status Tracker} tracks the force evaluation of cells with one entry tracks per cell. To start motion update of a certain cell, all its particles, and of its 26 neighborcells, must be fully evaluated. When that happens, the entries tracking the cell, as well as its neighborcells, are shifted right 1 step. The initial value of each entry is a one followed by 27 zeros. Once a cell and its 26 neighbors are all evaluated, the most-right bit of the corresponding entry becomes 1 and the scoreboard will send access request to the~\textit{Request Holder}. 

Since there is only one summation pipeline, summing particles/cells is sequential. The Request Holder is used to deal with the scenario when the force summation of a cell is still in progress, but access requests for other cells have been received. The Request Holder sends access requests to the address generator using round-robin. Once the address generator receives an access request, it can access the LR, RL, and Bonded Forces from the respective caches. The forces are summed and the results used for motion update.

\subsection{Motion Update and Particle Migration}

\begin{figure}[ht]
\centering
    \centering
    \includegraphics[width=0.45\textwidth]{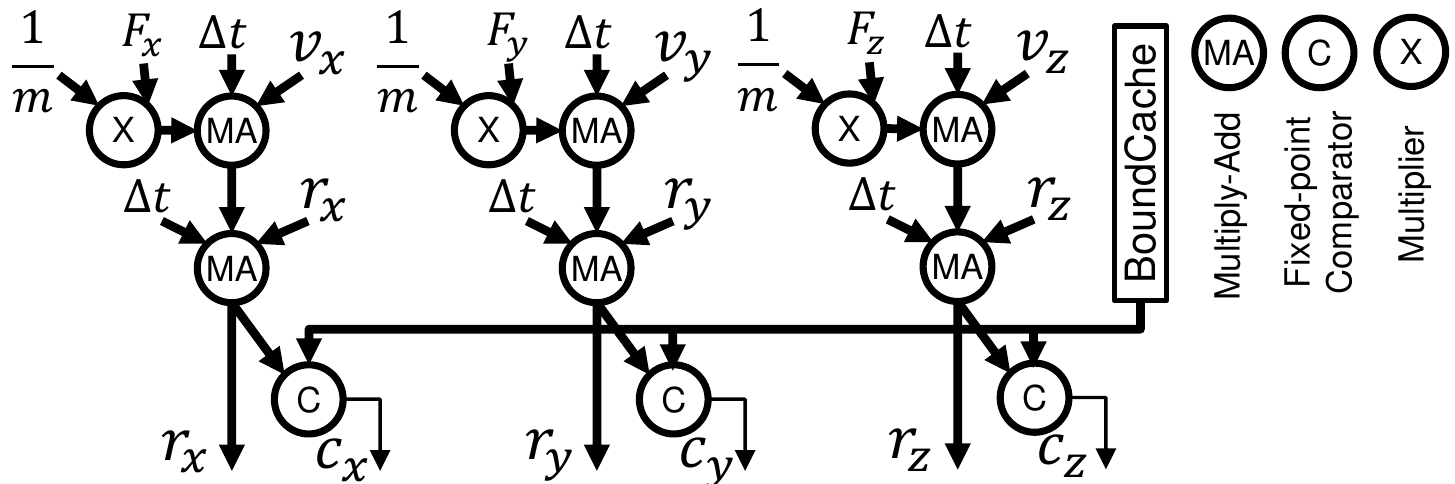}
\vspace*{-0.1truein}
    \captionof{figure}{Motion Update Pipeline}
    \label{fig:motion_update}
\end{figure}

Figure~\ref{fig:motion_update} shows the workflow inside the motion update module  following~\Cref{eq:motion_update_1,eq:motion_update_2,eq:motion_update_3}.
When the updated positions are calculated, the module computes the target cell of the updated particle. Since we are using a short timestep, and we perform motion integration each iteration, particles rarely move across more than one cell. Each cell has a pre-stored lower and upper boundary. If the updated position falls in the current cell range, the output cell id remain the same as the input. Otherwise, the output cell id will add or subtract one depending on the comparison with the boundary value.


\subsection{Workload-aware Adoption}
\label{ssec:workload_adoption}

There are numerous places where the design can be optimized with respect to workload. 
So that a user does not have to be concerned with this, we have implemented a workload-aware hardware generator script. Given the dataset size and hardware resources, it will provide a coarse estimate of resource usage and simulation performance for various mapping schemes. This script, along with the scripts used for generating data for particle cache, interpolation indices, cell boundaries, etc., will be made publicly available.

\section{Evaluation}

\subsection{Experimental Setup}

We have implemented, tested, and verified our designs on a Reflex XpressGX S10-FH200G Board which hosts an Intel Stratix 10 1SG280LU2F50E2VG chip \cite{Reflex18}. This chip is high-end with: 933,120 ALMs, 11,721 RAM blocks, and 5,760 DSP units, which makes it a good target for implementing FPGA/MD. 
To get the comparable MD simulation performance on CPU and GPU, we installed Amber 18~\cite{Case18} on a single node with an Intel Xeon Skylake CPU and various Nvidia GPUs. 

The dataset we use is Dihydrofolate Reductase (DFHR), a 159-residue protein in water, with 23,588 atoms~\cite{Case18}. The dataset is constrained to a bounding box of 62.23$\times$62.23$\times$62.23\AA, with a cutoff radius of 9\AA. The simulation timestep if 2$fs$ with Particle Mesh Ewald (PME) every two iterations.



\subsection{RL Performance Trade-offs}

\subsubsection{Mux-tree Resource Usage}
In Section~\ref{sssec:mux_tree}, we proposed a tree-based one-to-all mux to boost the operating frequency when broadcast position data from a single position memory module to multiple force pipelines. In this part, we will evaluate the cost for implementing such a mux tree (1-to-100) and comparing with a naive implementation of having single wire driving the input on 100 pipelines. The resource utilization is shown in Table~\ref{tab:mux_tree_usage}. We design our force pipeline that it takes one pair of particle position each cycle running at 350MHz. Thus in order to avoid pipeline starving, we have chosen the mux tree design, even though it has high resource utilization.

\begin{table}[ht]
\vspace*{-0.2truein}
\caption{Mux tree resource usage (size: 1-to-100)}
\vspace*{-0.1truein}
    \label{tab:mux_tree_usage}
    \centering
    \begin{tabular}{|c||c|c|c|}
        \hline
          Design                & ALM         & Frequency(MHz)  & Latency         \\    \hline
          Direct Connection     & 190         & 98              & 1     \\    \hline
          Mux Tree              & 5623        & 421             & 3     \\    \hline
    \end{tabular}
\vspace*{-0.2truein}
\end{table}

\subsubsection{RL Filter Resource Usage}

In Section~\ref{ssec:filter_implemetation} we propose two designs. Since the planar method requires an extra datapath in the force pipeline to generate $r^2$, we evaluate aggregate resource usage, including both filter bank and force pipeline. Table~\ref{tab:filter_usage} gives the resource usage of a single bank consisting of a force evaluation unit and eight filters. We note that a 10\% increase in ALM usage saves 74\% on DSPs; the planar method thus enables more pipelines per FPGA. All following evaluations assume planar filters.

\begin{table}[h]
\caption{Filter bank resource usage under two implementations}
\vspace*{-0.1truein}
    \label{tab:filter_usage}
    \centering
    \begin{tabular}{|c||c|c|c|}
        \hline
          \backslashbox{Design}{Usage}    & ALM          &  BRAM      & DSP         \\    \hline
          Direct Computation            & 5077(0.5\%) & 66(0.6\%) & 57(1\%)     \\    \hline
          Planar                        & 5605(0.5\%) & 65(0.6\%) & 15(0.3\%)     \\    \hline
    \end{tabular}
\vspace*{-0.1truein}
\end{table}

\subsubsection{RL Force Pipeline Comparison}
\label{sssec:force_pipe_resource}
 Section~\ref{ssec:force_eval} describes decisions between fixed and float and direct and interpolated. Resource utilization is shown in Table~\ref{tab:pipeline_usage}. We use first order interpolation with 256 intervals. Again, the deciding factor is DSP usage, which favors interpolation with floating-point.

\begin{table*}[h]
\caption{Force evaluation pipeline resource usage and performance comparisons}
\vspace*{-0.15truein}
    \label{tab:pipeline_usage}
    \centering
    \begin{tabular}{|c|c||c|c|c||c|c|}
        \hline
          Design            & Datatype              & ALM           &  BRAM         & DSP       & Frequency (MHz)   & Latency(Clock Cycle)\\    \hline
          Direct Computation& 32-bit Fixed-Point    & 1365(0.15\%)  & 4(0.04\%)     & 19(0.33\%)& 505               & 89\\    \hline
          Interpolation     & 32-bit Fixed-Point    & 866(0.09\%)   & 17(0.15\%)    & 10(0.17\%)& 433               & 22\\    \hline
          Direct Computation& Single Float          & 698(0.08\%)   & 3(0.03\%)     & 11(0.19\%)& 486               & 59\\    \hline
          Interpolation     & Single Float          & 462(0.05\%)   & 17(0.15\%)    & 6(0.09\%) & 654               & 14\\    \hline
    \end{tabular}
\end{table*}

\subsection{LR Performance Trade-offs}

We parameterize the LR modules to make trade-offs between LR evaluation time and resource usage. Here we specifically show the effect of varying particle to grid mapping modules, while maintaining a constant number of input particles (see Table~\ref{tab:LR_performance}). We note overall that the mapping units have a small effect on resource consumption. For performance, however, adding a single mapping unit improves performance by more than a third. Beyond this, however, the FFT \& IFFT latency becoming the dominant factor. 

\begin{table}[!htbp]
\caption{LR resource usage and evaluation time}
\vspace*{-0.15truein}
    \label{tab:LR_performance}
    \centering
    \begin{tabular}{|c||c|c|c|c|c|}
        \hline
         \thead{\# LR\\GridMapping Units}    & ALM           &  BRAM         & DSP       & \thead{Latency\\(Clock Cycle)}\\    \hline
          1                                 & 209,284       & 2,608         & 1,845     & 190,069                       \\    \hline
          2                                 & 210,406       & 2,608         & 2,019     & 119,395                       \\    \hline
          3                                 & 211,528       & 2,608         & 2,193     & 106,307                       \\    \hline
          4                                 & 212,650       & 2,608         & 2,367     & 101,727                       \\    \hline
    \end{tabular}
\end{table}


\subsection{Bonded Force Performance Trade-offs}
\label{ssec:bond_performance}
In Section~\ref{ssec:bonded_pipeline}, we propose merging three bonded force pipelines into a single one. Table~\ref{tab:bonded_usage} shows the benefits of this approach. 
The proposed merged pipeline saves 27\%, 43\%, 25\% on ALM, BRAM, and DSP, respectively. 
As the bonded force is still almost twice as fast as LR and RL this design decision is justified.

\begin{table}[h]
\caption{Bonded force pipeline resource usage}
\vspace*{-0.15truein}
    \label{tab:bonded_usage}
    \centering
    \begin{tabular}{|c||c|c|c|c|c|}
        \hline
                    & ALM       & BRAM    & DSP     & \thead{Latency\\(Clock Cycles)}  & \thead{Frequency\\(MHz)} \\    \hline
          Bond      & 1,481     & 10      & 18      & 148      & 398     \\    \hline
          Angle     & 13,691    & 77      & 153     & 187      & 401     \\    \hline
          Dihedral  & 12,432    & 77      & 201     & 244      & 392     \\    \hline
          Merged    & 20,109    & 93      & 278     & 276      & 330     \\    \hline
    \end{tabular}
\end{table}


\subsection{Full System Performance}

\subsubsection{Overall System Resource Utilization}

\begin{table*}[h]
\caption{Full system resource usage. Columns 2-4 are post place\&route. Columns 5-6 give the number of replications of RL pipeline and LR Grid Mapping units in each design. The last two give the stand-alone performance of RL and LR units.}
\vspace*{-0.1truein}
    \label{tab:total_usage}
    \centering
    \begin{tabular}{|c||c|c|c|c|c|c|c|}
        \hline
          Design                      & ALM              &  BRAM         & DSP            & \# RL Pipes     & \thead{\# LR\\GridMapping Units}   & \thead{RL Iter Time\\($\mu$s)}    & \thead{LR Iter Time\\($\mu$s)}  \\    \hline
          Design 1: Mem 1 + Dis 1     & 657,808(70.5\%)  & 9,430(80.4\%) & 4,419(76.7\%)  & 52              & 1                                 & 64,376.87                         &  817.56  \\    \hline
          Design 2: Mem 2 + Dis 1     & 747,075(80.0\%)  & 9,077(77.4\%) & 4,338(75.4\%)  & 35              & 2                                 & 349.30                            &  513.45  \\    \hline
          Design 3: Mem 1 + Dis 2     & 657,508(70.5\%)  & 9,430(80.5\%) & 4,038(70.1\%)  & 52              & 1                                 & 968.95                            &  817.56  \\    \hline
          Design 4: Mem 2 + Dis 2     & 746,775(80.0\%)  & 9,077(77.4\%) & 3,957(68.7\%)  & 35              & 2                                 & 292.89                            &  513.45  \\    \hline
          Design 5: Mem 1 + Dis 3     & 646,946(69.3\%)  & 9,362(79.9\%) & 4,197(72.8\%)  & 51              & 2                                 & 270.72                            &  513.45  \\    \hline
          Design 6: Mem 2 + Dis 3     & 586,336(62.8\%)  & 9,362(79.9\%) & 4,047(70.3\%)  & 41              & 2                                 & 260.37                            &  513.45  \\    \hline
    \end{tabular}
\end{table*}

As described in Section~\ref{ssec:LR_imple}, RL, LR, and Bonded are designed for balanced load.
To recap, as introduced in Section~\ref{sssec:workload_distribution}, we have two particle-to-memory mapping schemes: mapping all the particle in a single large memory unit and mapping particles onto small block RAMs based on the cell it belongs to. We also have three workload to pipeline mapping schemes: all pipelines work on same reference particle, all pipelines work on the same homecell with different reference particles, and each pipeline works on a different homecell. This yields six different designs. 

Table~\ref{tab:total_usage} lists the resource utilization and the number of function units that can fit onto a single FPGA-chip under different RL mapping schemes. We also list the stand-alone performance number for both RL and LR parts. By adjusting the number of \textit{LR Particle to Grid Mapping} modules (column 6), we aim to make the LR evaluation time about twice as much as RL (column 7 \& 8). 

We note first that Designs 2 \& 4 can only fit 35 pipelines. Those two designs have hundreds of memory modules, while the workload mapping requires each pipeline to receive data from all cells. Because of this, a very large on-chip switch (mux-tree based) is required, which consumes a large number of ALMs (190K). Compared with \textit{Mem 2}, designs using \textit{Mem 1} all have more pipelines, due to the convenience of having a single source of input data. Given the resource usage comparison, it seems that having a global memory provides benefits of having more pipelines mapped on to a single chip. However, the stand-along RL performance shows otherwise. We describe this next.

\subsubsection{MD System Performance}

Table~\ref{tab:performance} lists performance numbers for the DHFR dataset on various platforms, including multi-core CPU, GPU, and our six HDL implementations on different FPGAs. The CPU and GPU numbers come from collaborators in an industrial drug design environment. Compared with the best-case single CPU performance, the best-case FPGA design has one order of magnitude better performance. The FPGA design has 10\% more throughput than that of the GPU performance. Much more evaluation needs to be done, but we believe these results to be promising.

As shown in Table~\ref{tab:total_usage}, RL is the limiting factor on the overall performance. The poor performance of Design 1 is due to the memory bandwidth limitation:
for most cycles, pipelines are waiting for data. In Design 2, the distributed memory provides much higher read bandwidth. Design 3 faces a different problem: the number of particles per cell (70) is not a multiple of the number of pipelines (52), which means a set of pipelines (18) is idle after evaluating a single reference particle. It also suffers from memory bandwidth limitations. Design 4 has a happy coincidence that its pipeline count (35) can be divided evenly into 70 and most pipelines will have close to 100\% usage (this is subject to dataset). Designs 5 \& 6 might be supposed to have similar performance, but in Design 5 there is overhead on reading the first sets of input data from a single memory unit. But the subsequent read latency can be fully hidden.


\begin{table}[h]
\caption{Performance comparison: the middle column shows time to perform one full iteration (22k dataset); the right column shows throughput with a 2$fs$ timestep.}
\vspace*{-0.1truein}
    \label{tab:performance}
    \centering
    \begin{tabular}{|c||c|c|}
        \hline
          Platform          & \thead{Iteration Time\\($\mu$s)}  & \thead{Simulation Time\\(ns/day)}\\    \hline
          CPU 1-core        & 85,544                            & 2.02        \\    \hline
          CPU 2-core        & 38,831                            & 4.45        \\    \hline
          CPU 4-core        & 21,228                            & 8.14        \\    \hline
          CPU 8-core        & 11,942                            & 14.47       \\    \hline
          CPU 16-core       & 6,926                             & 24.95       \\    \hline
          GTX 1080 GPU      & 720                               & 240.13      \\    \hline
          Titan XP GPU      & 542                               & 318.97      \\    \hline
          RTX 2080Ti GPU    & 389                               & 444.05~\cite{Case18}      \\    \hline
          Design 1: Mem 1 + Distribution 1  & 64,411            & 2.68        \\    \hline
          Design 2: Mem 2 + Distribution 1  & 370               & 467.40      \\    \hline
          Design 3: Mem 1 + Distribution 2  & 1003              & 172.36      \\    \hline
          Design 4: Mem 2 + Distribution 2  & 313               & 551.55      \\    \hline
          Design 5: Mem 1 + Distribution 3  & 291               & 593.55      \\    \hline
          Design 6: Mem 2 + Distribution 3  & 274               & 630.25      \\    \hline
    \end{tabular}
\end{table}


\subsection{Dataset Impact on Mapping Selection}

Our system takes advantage of FPGAs' reconfigurability to fully customize the number of pipelines and the mapping scheme of workload and particle storage. Since RL evaluation takes both most of the resources and evaluation time, we focus here on examining the RL performance. We provide a software script that can quickly estimate the number of pipelines and resource usage based on the size of the input dataset and number of cells, along with an estimation of the simulation performance from the six different mapping schemes. In order to further demonstrate the selection of mapping schemes, we use a variety of datasets (5K to 50K) and cutoff radii (leading to different cell sizes). Characteristics are shown in Table~\ref{tab:datasets}.

\begin{table}[h]
\caption{Various testing datasets evaluating the impacts on workload mapping selection}
\vspace*{-0.1truein}
    \label{tab:datasets}
    \centering
    \begin{tabular}{|c||c|c|c|}
        \hline
                        & Particle \#   & Cell \#       & Particle \#/Cell  \\    \hline
          Dataset 1     & 5,000         & 63            & 80                \\    \hline
          Dataset 2     & 5,000         & 12            & 417               \\    \hline
          Dataset 3     & 20,000        & 252           & 80                \\    \hline
          Dataset 4     & 20,000        & 50            & 400               \\    \hline
          Dataset 5     & 50,000        & 625           & 80                \\    \hline
          Dataset 6     & 50,000        & 125           & 400               \\    \hline
    \end{tabular}
\end{table}

\begin{figure}[ht]
\centering
    \centering
    \includegraphics[width=0.47\textwidth]{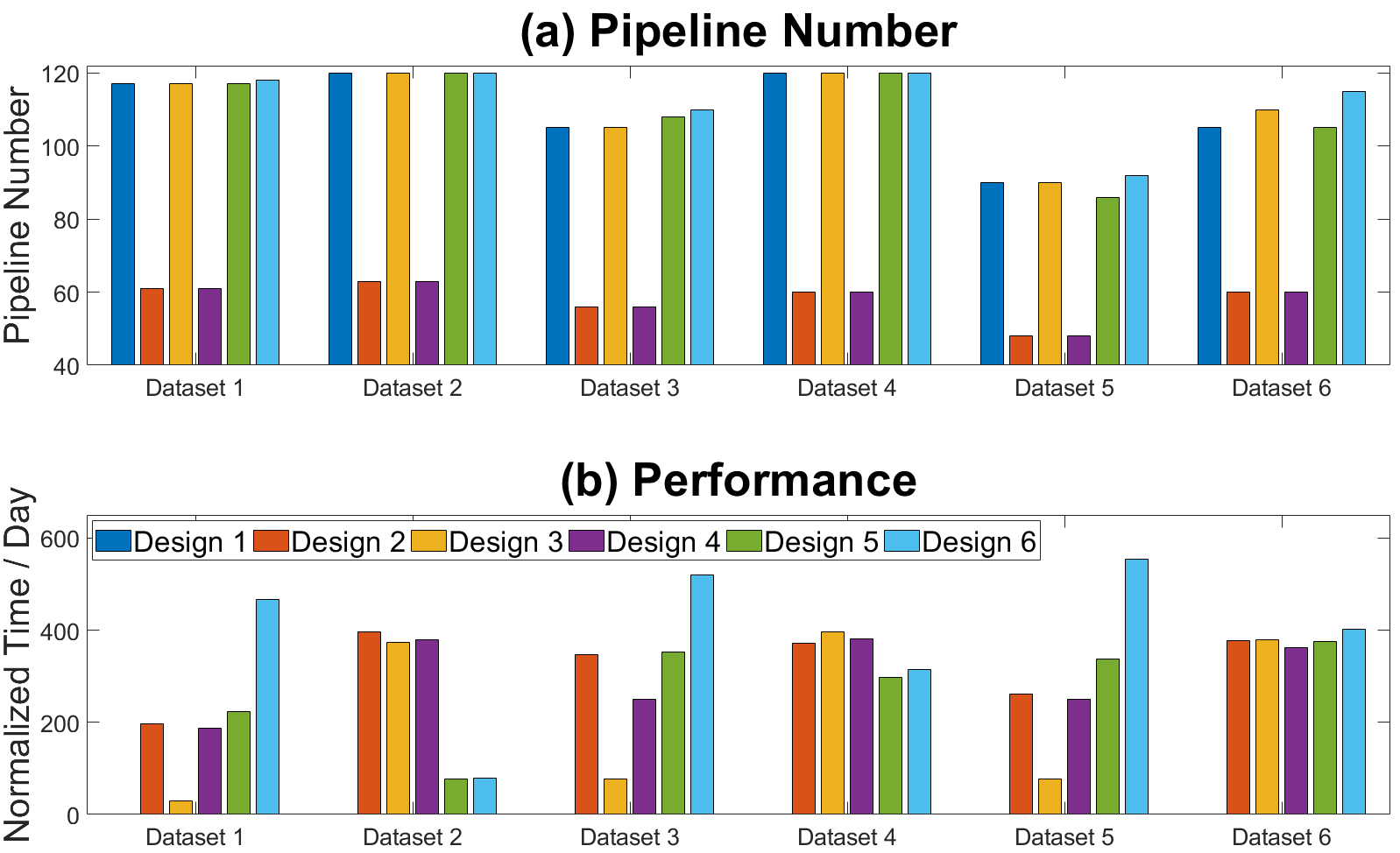}
    \captionof{figure}{Performance with Different Datasets: (a) Number of RL pipelines that can map onto a single FPGA; (b) RL simulation performance, normalized to Design 1 for each dataset.}
    \label{fig:performance_vs_datasets}
\end{figure}

The number of pipelines and performance are shown in Figure~\ref{fig:performance_vs_datasets}. We note first (from Figure~\ref{fig:performance_vs_datasets}a) that the dataset size has little impact on the number of pipelines we can map on a single Stratix 10 FPGA until the dataset grows large enough to cause a resource conflict (in BRAMs). However, this is not the case on simulation performance as shown in Figure~\ref{fig:performance_vs_datasets}b. All the performance number is normalized to the Design 1 performance for each dataset. We have the following observations: \textbf{(i)} Design 1 with single particle memory and workload distribution 1 always has the worst performance due to memory bottleneck; \textbf{(ii)} When the dataset is sparse (see Dataset 1, 3, 5), Design 6 tends to return the best performance, and the relative performance among the six designs is similar; \textbf{(iii)} When the dataset is dense (see Dataset 2, 4, 6), workload distribution 3 provides fewer benefits comparing with workload distribution 2; this is especially clear when the dataset is small and dense. 

\section{Summary and Future Work}
\label{sec:conclusion}

We present an end-to-end MD system on a single FPGA featuring online particle-pair generation, force evaluation on RL, LR, and bonded interactions, motion update, and particle data migration. We provide an analysis of the most likely mappings among particles/cells, BRAMs, and on-chip compute units. We introduce various microarchitecture contributions on routing the accumulation of hundreds of particles simultaneously and integrating motion update. A set of software scripts is created to estimate the performance of various design choices based on different input datasets. 
We evaluate the single-chip design on a commercially available Intel Stratix 10 FPGA and achieve a simulation throughput of 630ns/day on a 23.5K DFHR dataset, which is at least 40\% better than the analogous state-of-the-art GPU implementations.


%

%
\bibliographystyle{unsrt}
\bibliography{comm_refs,d_refs,caad_refs,fpga_eda_refs,md_refs,MPI,bcb_refs}

\end{document}